\documentclass[fleqn,usenatbib]{mnras}

\usepackage[T1]{fontenc}

\DeclareRobustCommand{\VAN}[3]{#2}
\let\VANthebibliography\thebibliography
\def\thebibliography{\DeclareRobustCommand{\VAN}[3]{##3}\VANthebibliography}

\def\equationautorefname#1#2\null{(#2)}

\usepackage{graphicx}	 % Including figure files
\usepackage{amsmath}	 % Advanced maths commands
\usepackage{amssymb}	 % Extra maths symbols
\usepackage{mathtools}   % multi-lines for equation
\usepackage{CJK}         % Chinese name
\usepackage{xcolor}      % comments color
\usepackage{scalerel}
\usepackage{tikz}
\usetikzlibrary{svg.path}
\definecolor{orcidlogocol}{HTML}{A6CE39}
\tikzset{
  orcidlogo/.pic={
    \fill[orcidlogocol] svg{M256,128c0,70.7-57.3,128-128,128C57.3,256,0,198.7,0,128C0,57.3,57.3,0,128,0C198.7,0,256,57.3,256,128z};
    \fill[white] svg{M86.3,186.2H70.9V79.1h15.4v48.4V186.2z}
                 svg{M108.9,79.1h41.6c39.6,0,57,28.3,57,53.6c0,27.5-21.5,53.6-56.8,53.6h-41.8V79.1z M124.3,172.4h24.5c34.9,0,42.9-26.5,42.9-39.7c0-21.5-13.7-39.7-43.7-39.7h-23.7V172.4z}
                 svg{M88.7,56.8c0,5.5-4.5,10.1-10.1,10.1c-5.6,0-10.1-4.6-10.1-10.1c0-5.6,4.5-10.1,10.1-10.1C84.2,46.7,88.7,51.3,88.7,56.8z};
  }
}

\usepackage{newtxtext,newtxmath}

\newcommand\orcidicon[1]{\href{https://orcid.org/#1}{\mbox{\scalerel*{
\begin{tikzpicture}[yscale=-1,transform shape]
\pic{orcidlogo};
\end{tikzpicture}s
}{|}}}}

\title[Dynamical Effects of Magnetic Opacity]{Dynamical Effects of Magnetic Opacity in Neutron Star Accretion Columns}

\author[X. Sheng et al.]{
Xin Sheng (盛昕) \orcidicon{0009-0007-2375-2327},$^{1}$\thanks{Email: xs2525@columbia.edu}
Lizhong Zhang (张力中) \orcidicon{0000-0003-0232-0879},$^{1}$
Omer Blaes \orcidicon{0000-0002-8082-4573},$^{1}$
Yan-Fei Jiang (姜燕飞) \orcidicon{0000-0002-2624-3399}$^{2}$
\\
$^{1}$Department of Physics, University of California, Santa Barbara, CA 93106, USA\\
$^{2}$Center for Computational Astrophysics, Flatiron Institute, New York, NY 10010, USA\\
}

\date{Accepted XXX. Received YYY; in original form ZZZ}

\pubyear{2021}

\begin{document}
\begin{CJK*}{UTF8}{gbsn}
\label{firstpage}
\pagerange{\pageref{firstpage}--\pageref{lastpage}}
\maketitle

\begin{abstract}
We present relativistic, radiation magnetohydrodynamic simulations of supercritical neutron star accretion columns in Cartesian geometry, including temperature-dependent, polarization-averaged Rosseland mean opacities accounting for classical electron scattering in a magnetic field.  Just as in our previous pure Thomson scattering simulations, vertical oscillations of the accretion shock and horizontally propagating entropy waves (photon bubbles) are present in all our simulations.  However, at high magnetic fields $\gtrsim10^{12}$~G, the magnetic opacities produce significant differences in the overall structure and dynamics of the column.  At fixed accretion rate, increasing the magnetic field strength results in a shorter accretion column, despite the fact that the overall opacity within the column is larger.  Moreover, the vertical oscillation amplitude of the column is reduced.  Increasing the accretion rate at high magnetic fields restores the height of the column.
However, a new, slower instability takes place at these field strengths because they are in a regime where the opacity increases with temperature.  This instability causes both the average height of the column and the oscillation amplitude to substantially increase on a time scale of $\sim10$~ms.  We provide physical explanations for these results, and discuss their implications for the observed properties of these columns, including mixed fan-beam/pencil-beam emission patterns caused by the oscillations.
\end{abstract}


%%%%%%%%%%%%%%%%%%%%%%%%%%%%%%%%%%%%%%%%%%%%%%%%%%

%%%%%%%%%%%%%%%%% BODY OF PAPER %%%%%%%%%%%%%%%%%%

\section{Introduction}

Accretion of matter onto a magnetized compact object is an important process in a variety of astronomical contexts, from polars and intermediate polars in white dwarf binaries to accretion-powered pulsars in neutron star binaries.  Inside the Alfv\'en radius, where magnetic stresses are comparable in magnitude to the ram pressure of the accreting material, matter is thought to be magnetically guided through the compact object's magnetosphere toward the magnetic poles \citep{ghosh1977}.  In many systems the material lands on the stellar surface, resulting in hot spots that locally emit more or less isotropically (so-called pencil-beam emission).  However, in high mass X-ray binary neutron star systems where the local accretion rate can be a significant fraction of Eddington, radiation pressure can result in an accretion shock above the stellar surface, below which a quasi-hydrostatic region forms in which matter subsonically settles down onto the stellar surface.  This accretion column structure radiates much of the accretion luminosity from its sides \citep{Inoue1975,basko1976} in what has come to be known as fan-beam emission.

Accretion columns are central to all models of high luminosity X-ray pulsars and pulsating ultra-luminous X-ray sources (see \citealt{mus22,king2023} for recent reviews).  These models generally assume a simplified static, 1D structure, but in fact these columns are theoretically expected to have significant time-dependent behavior with substantial lateral spatial complexity \citep{arons1987}. Multidimensional numerical simulations are an essential tool to elucidate this behavior \citep{klein1996,kawashima2020}.

This is the fourth in a series of papers on radiation magnetohydrodynamic
simulations of neutron star accretion columns using the code \textsc{Athena++} \citep{STO20} with the radiation module developed by \citet{Jiang2021}.  Previous papers in this series examined the nonlinear development of photon bubble instabilities in static neutron star atmospheres \citep{Paper1}, the nonlinear dynamics of short accretion columns in Cartesian geometry \citep{Paper2}, and the nonlinear dynamics of more global, large accretion columns in split monopole magnetic fields \citep{Paper3}.  All the previous simulations in this series assumed classical, isotropic Thomson scattering for the Rosseland mean opacity, and did not account for any magnetic effects on this opacity.  Indeed, the only simulations that we are aware of that attempted to account for such magnetic opacity effects were by \citet{klein1996}.

For the high magnetic fields that are typical of young neutron stars, the magnetic field can significantly affect the opacity.  This can greatly reduce the radiation pressure force on the plasma at low temperatures (e.g. \citealt{canuto1971,lodenquai1974,meszaros1980,arons1987}), and even increase the radiation pressure force for temperatures comparable to the cyclotron energy \citep{Suleimanov2022}.  This can dramatically affect the dynamics of the accretion column, and possibly even introduce new instabilities.  In this paper, we simulate these effects for columns with different magnetic fields and accretion rates, exploring in particular their variability and resulting light curves.

This paper is organized as follows.  In \autoref{sec:numerical_method}, we describe the numerical methods that we employ in the simulations.  In \autoref{sec:results}, we present our simulation results, discuss the physics that drives their behavior, and discuss the properties of the emergent radiation. In \autoref{sec:discussion}, we discuss some of the numerical caveats and also mention some observationally testable predictions.  We then summarize our results in \autoref{sec:conclusions}.

\section{Numerical Method}
\label{sec:numerical_method}

\subsection{Equations}
\label{sec:equations} 

\begin{figure}
  \centering
  \includegraphics[width=\columnwidth]{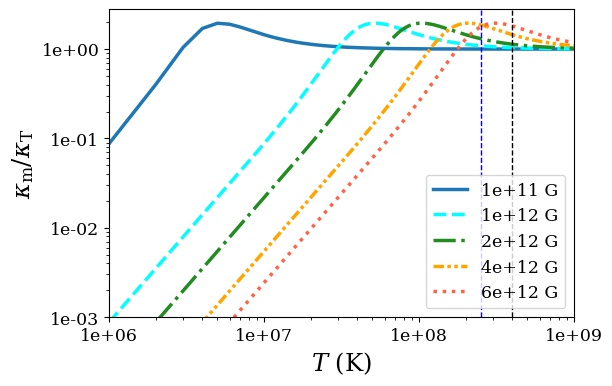}
  \caption{Temperature-dependence of the magnetic opacities that we use in our simulations, in units of the Thomson opacity.  Each curve represents the different magnetic fields that we adopt for our simulations. The vertical dashed lines indicate the approximate maximum temperatures achieved at the base of the column in our simulations:  $\sim2.5\times10^8$~K for the low accretion rate (Lowacc) simulations, and $\sim4\times10^8$~K in the high accretion rate (Highacc) simulations.}
  \label{fig:opacity}
\end{figure}

We follow the numerical treatments in \citet{Paper2} to solve the fluid conservation laws together with radiative transfer incorporating magnetic opacity.  The governing equations are summarized as follows: 
\begin{subequations}
\begin{align}
    &\partial_0(\rho u^0) + \partial_j(\rho u^j) = S_{\mathrm{gr}1}
    \quad, 
	\label{eq:particle_conserv}
	\\
	&\begin{multlined}[t]
	\partial_0(w u^0u^i - b^0b^i)
	\\
	+ \partial_j\left(w u^iu^j + \left(P_{\rm g}+ \frac{1}{2}b_{\nu}b^{\nu}\right)\delta^{ij} - b^ib^j \right) = S_{\mathrm{gr}2}^i - S_{r2}^i
    \ , 
	\end{multlined}
	\label{eq:mom_conserv}
	\\
	&\begin{multlined}[t]
	\partial_0\left[w u^0u^0 - \left(P_g+ \frac{1}{2}b_{\nu}b^{\nu}\right) - b^0b^0\right]
	\\
	\mkern150mu + \partial_j(w u^0u^j - b^0b^j) = S_{\mathrm{gr}3} - S_{r3}
    \quad, 
	\end{multlined}
	\label{eq:energy_conserv}
	\\
	&\partial_0I + n^j\partial_j I = \mathcal{L}^{-1}(\bar{S}_r)
    \quad, 
	\label{eq:rad_transfer}
\end{align}
\end{subequations}
where the gas density $\rho$ and gas pressure $P_{\rm g}$ are defined in the fluid rest frame.  The four-velocity is defined as $(u^0, u^i)=\Gamma(1, v^i)$, where $\Gamma=(1-v_j v^j)^{-1/2}$ is the Lorentz factor and the three-velocity $v^i$ is in units of the speed of light.  The quantity $I$ is the frequency-integrated radiation intensity and is a function of position and photon propagation direction $n^i$.  Note that Latin indices indicate the spatial components of a three-vector (i.e. here $i=1,2,3$ refer to $x,y,z$, respectively), and Greek indices refer to the time-spatial components of a four-vector.  Given the three-vector magnetic field $B^i$, its four-vector form $b^{\mu}=(b^0, b^i)$ and total enthalpy $w$ are given by: 
\begin{subequations}
\begin{align}
    &b^0 = u_jB^j,\quad b^i = \frac{1}{u^0}(B^i + b^0u^i)
    \quad, 
    \\
    &w = \rho + \frac{\gamma}{\gamma-1}P_g + b_{\nu} b^{\nu}
    \quad.
    \label{eq:total_enthalpy}
\end{align}
\end{subequations}
Here the adiabatic index $\gamma=5/3$ is assumed for the ideal gas.  The gravitational source terms $S_{\mathrm{gr}1}$, $S_{\mathrm{gr}2}^i$, and $S_{\mathrm{gr}3}$ are derived from general relativity in the weak field limit, and can be found in appendix~B of \citet{Paper1}.  In equation \autoref{eq:rad_transfer}, $\mathcal{L}^{-1}$ is the Lorentz boost operator from the comoving frame to the lab frame.  The source term of the radiative transport $\bar{S}_r$ is intrinsically defined in the comoving frame, while the radiation source terms $S_{r2}^i$ and $S_{r3}$ that exchange momentum and energy between gas and radiation are defined in the lab frame.  Both of their formulations can be found in equations~(4) of \citet{Paper2}, except here we adopt the magnetic opacity $\kappa_{\mathrm m}$ for photon-electron scattering.  The details of the magnetic opacity we use in our simulations can be found in \hyperref[sec:magnetic_opacity_derivation]{Appendix A}.  In our previous paper \citep{Paper2}, we assumed for simplicity that the scattering opacity was isotropic and constant and given by the Thomson value $\kappa_{\mathrm{T}}$.  However, this can significantly overestimate the photon-electron interaction when the magnetic field is strong at low gas temperatures.  Moreover, magnetic electron scattering is anisotropic and polarization dependent.  Nevertheless, a Rosseland and polarization-averaged opacity can be derived which is approximately isotropic (\citealt{arons1987}; \hyperref[sec:magnetic_opacity_derivation]{Appendix A}).  It is this magnetic opacity $\kappa_{\mathrm{m}}$ that we use to replace the Thomson opacity that we adopted in our previous simulations. \autoref{fig:opacity} shows the temperature dependence of this magnetic opacity for different magnetic field strengths.  The magnetic opacity significantly increases with gas temperature $T$ until it reaches a peak value of $1.95\kappa_{\mathrm{T}}$ at $kT=0.385$ times the electron cyclotron energy $\hbar\omega_{\rm ce}$.  It then returns to Thomson at higher temperatures.

\subsection{Simulation Setup}
\label{sec:simulation_setup} 

\subsubsection{Simulation Parameters}
For our simulations, we select version~2 (HR-Wide-25) in \citet{Paper2} as the prototype, and vary its magnetic field strength and accretion rate to explore the dynamical changes of the accretion column with magnetic opacity.  For all of our simulations, the mesh grids are set to be $700\times2048$ in the horizontal and vertical directions, respectively.  The domain height is $0.35 R_{\star}$, and the width of the accretion column region is $0.06 R_{\star}$, where we adopt a neutron star radius $R_{\star}=10^6$~cm.  

The global parameters that we vary in our seven simulations are listed in \autoref{tab:simparameters}.  For the first five simulations, we vary the magnetic fields that we use to compute the magnetic opacities, from $10^{11}$~G (Lowacc01) to $6\times10^{12}$~G (Lowacc6).  Simulations Lowacc1 ($10^{12}$~G) to Lowacc6 span the range of magnetic fields inferred from electron cyclotron line observations in X-ray pulsars \citep{sta19}, but we include Lowacc01 to compare with our previous Thomson scattering simulations.  For simulations Highacc4 and Highacc6, we increase the accretion rates for the simulations with the two strongest magnetic fields, in order to build up higher accretion columns.  The parameter $\epsilon$ is the accretion rate expressed as a local Eddington ratio and $\rho_{\mathrm{acc}}$ is the density of the incoming accretion flow at the top boundary.  Both of their definitions can be found in equation~(6) of \citet{Paper2}.

\begin{table}
	\centering
	\caption{Names and global parameters of all the simulations.  The magnetic fields listed are only used for the computation of the magnetic opacity.  The actual magnetic field used in the MHD was initialized to be $8\times10^{10}$~G for all the simulations. }
	\label{tab:example_table}
	\begin{tabular}{clccc} % four columns, alignment for each
		\hline
		Version & Name & Magnetic field & $\epsilon$ & $\rho_{\mathrm{acc}}$\\
            & & ($10^{12}$ G) & & ($10^{-4}$ g/cm$^3$)\\
		\hline
		1 & Lowacc01 & \,\,\,\,0.1 & \,\,\,25  & \,\,\,1.15 \\
		2 & Lowacc1 &         1   & \,\,\,25  & \,\,\,1.15 \\
		3 & Lowacc2 &         2   & \,\,\,25  & \,\,\,1.15 \\
        4 & Lowacc4 &         4   & \,\,\,25  & \,\,\,1.15 \\
        5 & Lowacc6 &         6   & \,\,\,25  & \,\,\,1.15 \\
        6 & Highacc4  &         4   &      375  &      17.25 \\
        7 & Highacc6  &         6   &      500  &      23.00 \\
		\hline
	\end{tabular}
	\label{tab:simparameters}
\end{table}

\subsubsection{Boundary and Initial Conditions}

The numerical setup of the simulation domain is identical to that of \citet{Paper2}, where the actual accretion column region is at the center with two vacuum regions on both sides and a gas-supported base at the bottom as effective boundaries.  The boundary conditions of the simulation domain are also the same as what we used in \citet{Paper2}, and we summarize them as follows.  The bottom boundary is reflective for both gas and radiation, and the magnetic fields are set to be constant and vertical. The side boundaries are reflective for both gas and magnetic fields, but allow the radiation to escape freely (i.e. a radiation vacuum boundary condition, see section~3.2 in \citealt{Paper1}).  Although the boundary conditions at the sides of the simulation domain are reflective for the magnetic field, the field at the edge of the actual accretion column inside the two vacuum regions is not so constrained.  The magnetic fields at the top boundary are set to be constant and vertical.  A cold accretion flow is injected from the top boundary within the accretion column region, where the comoving radiation fields are set to be isotropic and in local thermal equilibrium.  Outside the accretion column region, the top boundary is outflow for the gas and vacuum for the radiation so they are free to escape. 

For the five simulated accretion columns at the low accretion rate ($\epsilon=25$), we adopt as initial condition the 1D solution of the one-zone stationary model with Thomson opacity (for details, see section~2.3 of \citealt{Paper2}).  For the high accretion rate simulations, we start from the the corresponding low accretion rate simulation that uses the same magnetic field, select a snapshot when the accretion column is in the quasi-steady state, and restart it using the higher accretion rate.  To prevent numerical failures associated with a sudden accretion rate change, we slowly and linearly increase the accretion rate from $t=3000t_{\rm sim}$ to $t=4000t_{\rm sim}$, where the selected simulation time unit is $t_{\rm sim}=2.8\times10^{-7}$~s.

\subsection{Additional Numerical Treatments}
\label{subsec:additional_numerical_treatments}
The magnetic pressure in neutron star accretion columns is larger than the thermal pressure by orders of magnitude.  As a result, the variable inversion algorithm in \textsc{Athena++} in going from conservative to primitive variables (including gas pressure) can fail to %adequately
numerically resolve the small gas pressure.  This introduces numerical noise in determining the gas temperature which is used to define opacities and emissivities.  We therefore adopt an initial vertical magnetic field of $8\times10^{10}$~G for the actual magnetohydrodynamics (MHD) in all of the simulations presented in this paper. This is sufficiently strong to confine the column against horizontal radiation pressure forces, and to constrain the matter to move vertically. It is also low enough to avoid excessive numerical noise from the variable inversion in the sinking zone of the accretion column. Hence, the magnetic fields listed in \autoref{tab:simparameters} are purely used for the computation of the magnetic opacity, and not in the MHD itself.

However, the variable inversion algorithm can still fail in the low-density, free-fall region, which introduces substantial numerical noise into the gas temperature there.  Because the magnetic opacity depends on temperature, the opacity can also be noisy, and in fact sometimes achieves artificially high values when the temperature noise is substantial.  This can result in a strong interaction between the gas and radiation above the shock front.  In some of our early numerical experiments, this effect gradually destabilized the accretion column and eventually led to ejection of the incoming accretion flow above the shock front.  In reality, the matter in the free-fall zone has low temperature and so low magnetic opacity.  Therefore, in our simulations, we adopt a small, but nonzero, fixed value of the magnetic opacity ($\kappa_{\rm m} = 0.06\kappa_{\rm T}$) in the free-fall zone in order to eliminate this artificial noise and smoothly handle the transition between the free-fall zone and sinking zone.  

\section{Results}
\label{sec:results}

\begin{figure*}
    \centering
    \includegraphics[width=0.8\linewidth]{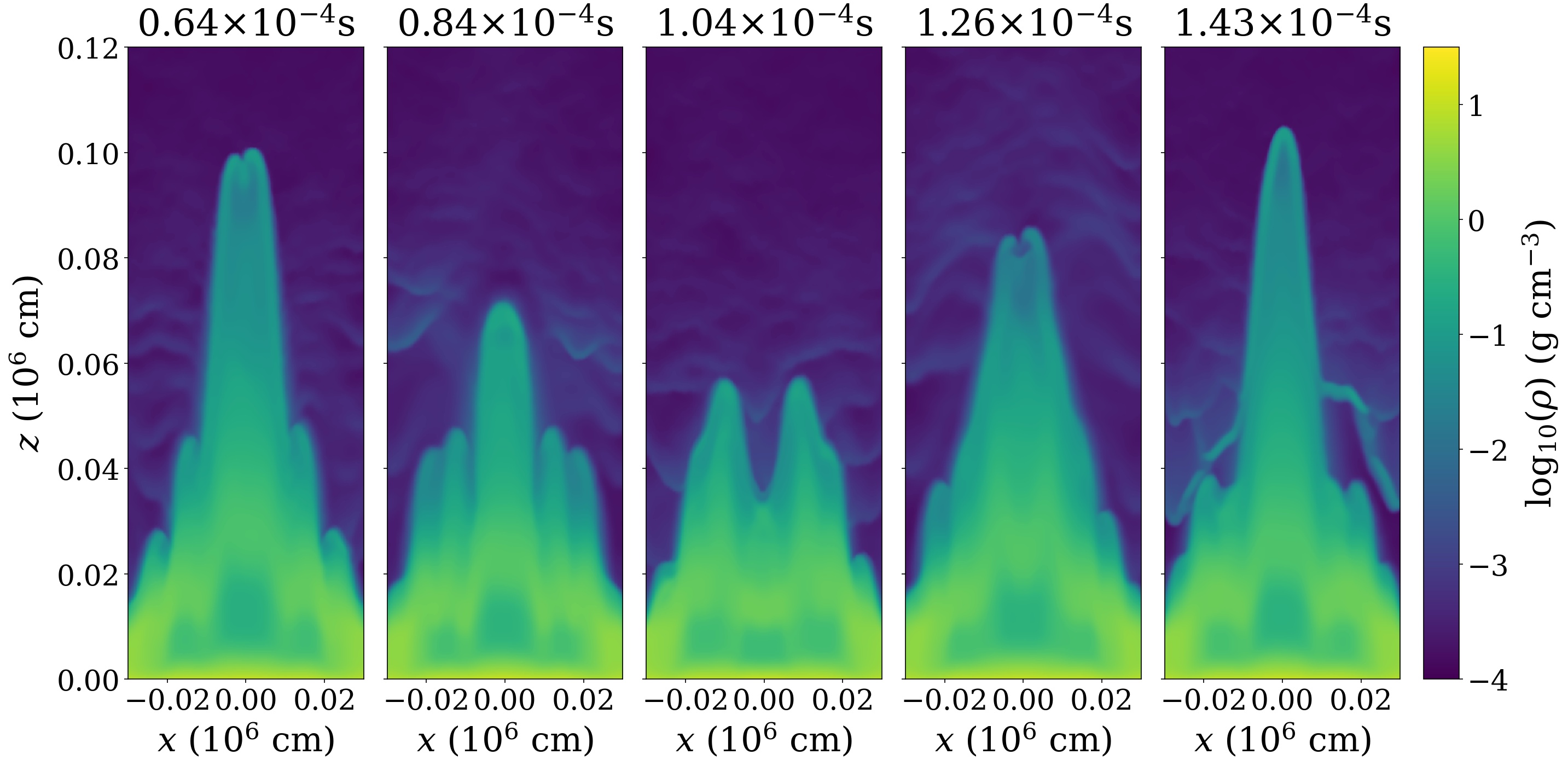}
    \caption{Snapshots of the density distribution in simulation Lowacc01, covering approximately one oscillation period of the shock at the center of the column.  (The times indicated in the panels correspond to the same origin as the time axis used to plot light curves in \autoref{fig:light_curve}.)
    }
    \label{fig:sinking_region_oscillation01}
\end{figure*}

\begin{figure*}
    \centering
    \includegraphics[width=0.8\linewidth]{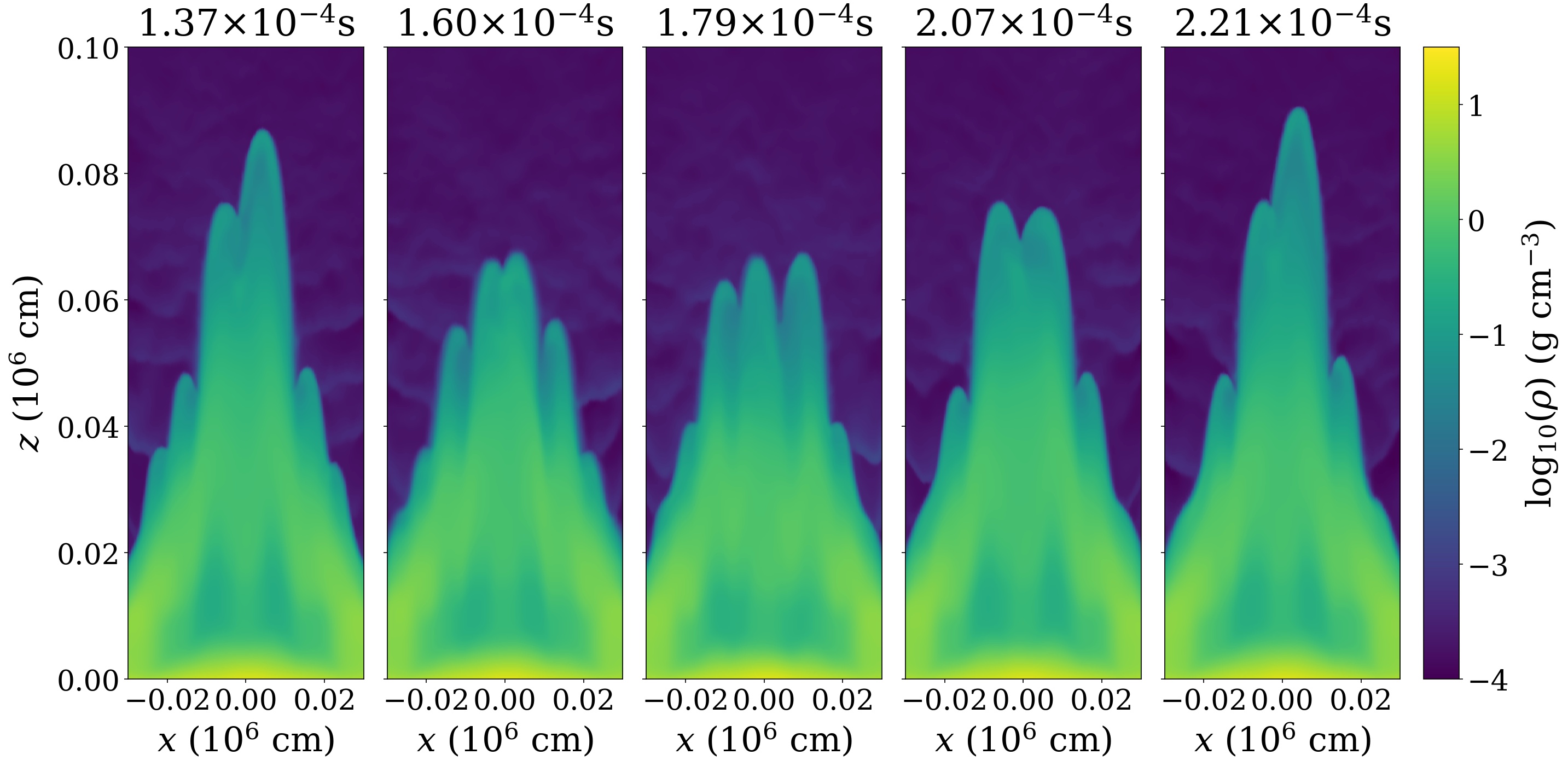}
    \includegraphics[width=0.8\linewidth]{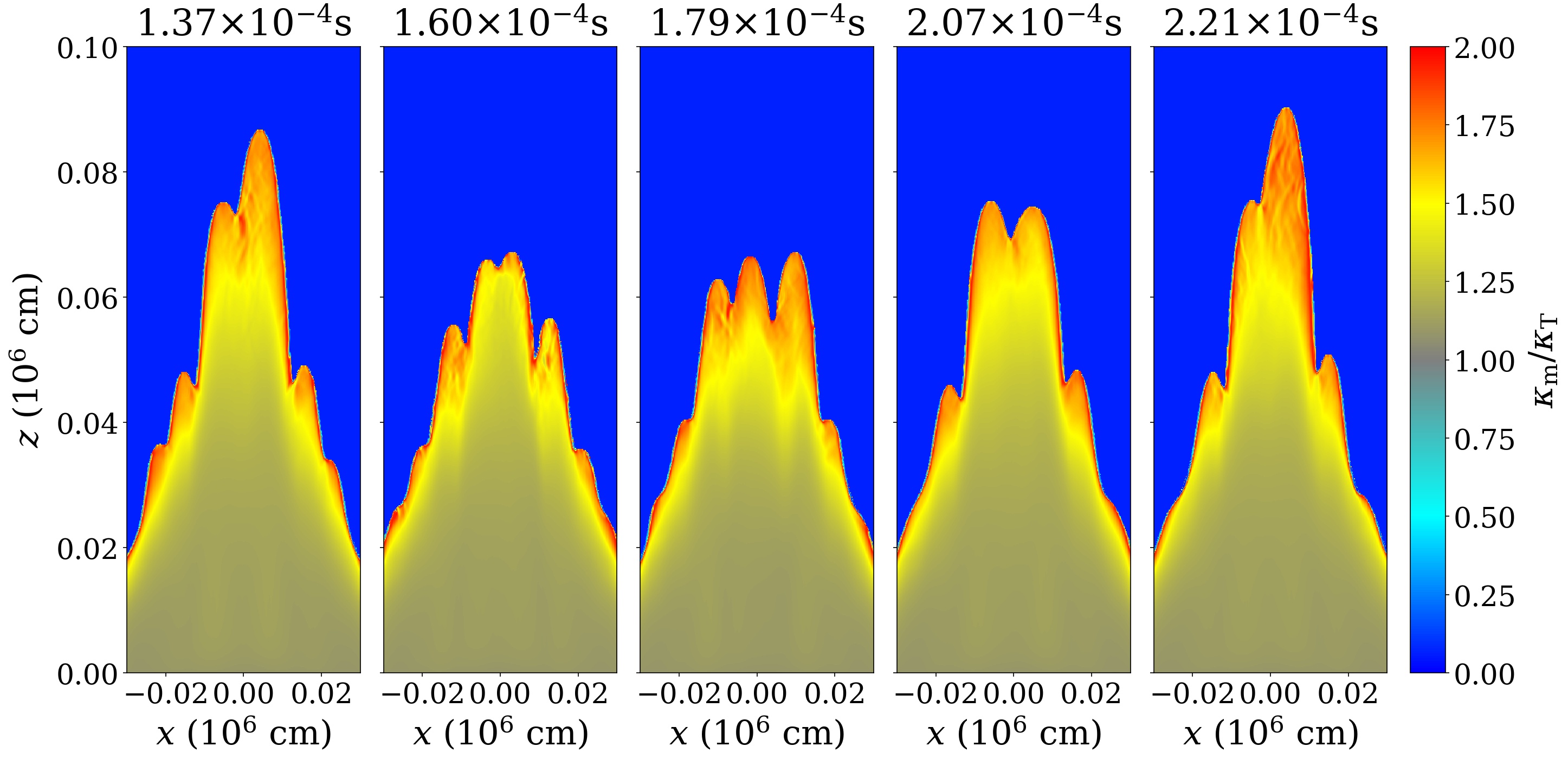}
    \caption{Snapshots of the density distribution in simulation Lowacc1, covering approximately one oscillation period.
    }
    \label{fig:sinking_region_oscillation1}
\end{figure*}

As noted above, our simulations with low accretion rates started from a 1D, Thomson scattering initial condition.  This always overestimates the column height because of (1) the underestimated cooling efficiency from the oversimplified top-hat column shape and (2) the altered radiation-gas interaction from magnetic opacity.  (We discuss the effects of this in detail in \autoref{subsec:Effects_from_the_Magnetic_Opacity}.) Therefore, the sinking zone quickly collapses and reaches a new equilibrium state where gravity is roughly balanced by the adjusted radiation pressure support.  After the accretion column has relaxed from the initial condition, the system gradually enters into a quasi-steady state with high-frequency oscillations that persist to the end of the simulation, similar to what we found in \citet{Paper2}.  For the simulations with high accretion rates, the column also reaches a quasi-steady state with oscillations.  However, at late times both the column height and the oscillation amplitude dramatically increase due to an instability associated with the fact that the opacity increases with temperature at these high magnetic fields (see \autoref{subsec:opacity_instability}).

\subsection{Behavior at Weak Magnetic Fields and Low Accretion Rate}
\label{subsec:Weak_Field_Low_Accretion_Rate}

In \autoref{fig:sinking_region_oscillation01}, we present the density distribution over one full oscillation period of Lowacc01 as an illustration of the oscillatory behavior. As discussed in \citet{Paper2}, the oscillation originates from the instantaneous mismatch between replenishment of internal energy and sideways radiative cooling.  When the accretion column is most vertically extended, the sideways radiative cooling is maximized, while the heat that is mostly generated at the shock front cannot be transported to the bottom fast enough to balance this cooling.  Hence, the column structure begins to collapse due to insufficient radiation pressure support.  When the accretion column is compressed to its lowest height, the sideways cooling is minimized and the low altitude shock front over-heats the column, resulting in vertical expansion.

\autoref{fig:sinking_region_oscillation01} also shows the presence of vertical finger-like structures that propagate horizontally inward toward the center of the column.  These are a manifestation of the entropy waves that are associated with the photon bubble instability in the slow radiation diffusion regime \citep{Paper2,Paper3}. These entropy waves are present in all of our simulations, but have little effect on the fundamental oscillatory dynamics, nor do they alter the oscillation frequency.

Compared with the analogous simulation using Thomson opacity (see HR-Wide-25 in \citealt{Paper2}), all the simulations using magnetic opacity lack the long-lived, coherent pre-shock structures because of the weak gas-radiation interaction when the gas temperature is far below the cyclotron energy in the free-fall zone. Nevertheless, as is evident from \autoref{fig:sinking_region_oscillation01}, density fluctuations do occur in the free-fall zone, particularly in the weaker magnetic field low accretion rate simulations.  As we discuss more extensively below in \autoref{subsec:Variability}, these simulations exhibit the strongest variability, and that variability is responsible for these fluctuations due to the interaction of the upward radiation field from the column.   The less variable simulations show significantly reduced fluctuations in the free-fall zone.  We also found that the pre-shock structure disappeared when the magnetic field decreased with height in the split monopole geometry of \citet{Paper3}, simply because of the reduced ram pressure at higher altitudes.

Lowacc01 has such a weak magnetic field that the temperature width of the opacity peak is easily surmounted in the shock and much of the sinking zone is supported by Thomson opacity.  (See the $10^{11}$~G curve in \autoref{fig:opacity}:  almost all of the temperature range toward the base of the column is in the Thomson regime).  It is therefore closest in its behavior to the simulations of our previous papers \citep{Paper2,Paper3} that assumed pure Thomson opacity.  \autoref{fig:sinking_region_oscillation1} depicts both the density and opacity in the stronger magnetic field simulation Lowacc1.  As illustrated by the $10^{12}$~G magnetic field curve in \autoref{fig:opacity}, the stronger magnetic field produces a wider opacity peak in temperature space.  This is evident in the lower panel of \autoref{fig:sinking_region_oscillation1}.  As material crosses the shock, the temperature climbs over the opacity peak and results in a high opacity in the post-shock plasma.  Going downward into the sinking zone, the opacity declines as we are past the opacity peak, eventually approaching the Thomson opacity in the deep interior.  Lowacc1 still shows substantial vertical shock oscillations and inward propagating entropy waves that were evident in Lowacc01.  As we will see in the next section, the effects of the opacity on the dynamics and structure of the column become much more important as we increase the magnetic field still further.

\subsection{Effects Arising from Changing Magnetic Field}
\label{subsec:Effects_from_the_Magnetic_Opacity}

\begin{figure*}
    \centering
    \includegraphics[width=0.8\linewidth]{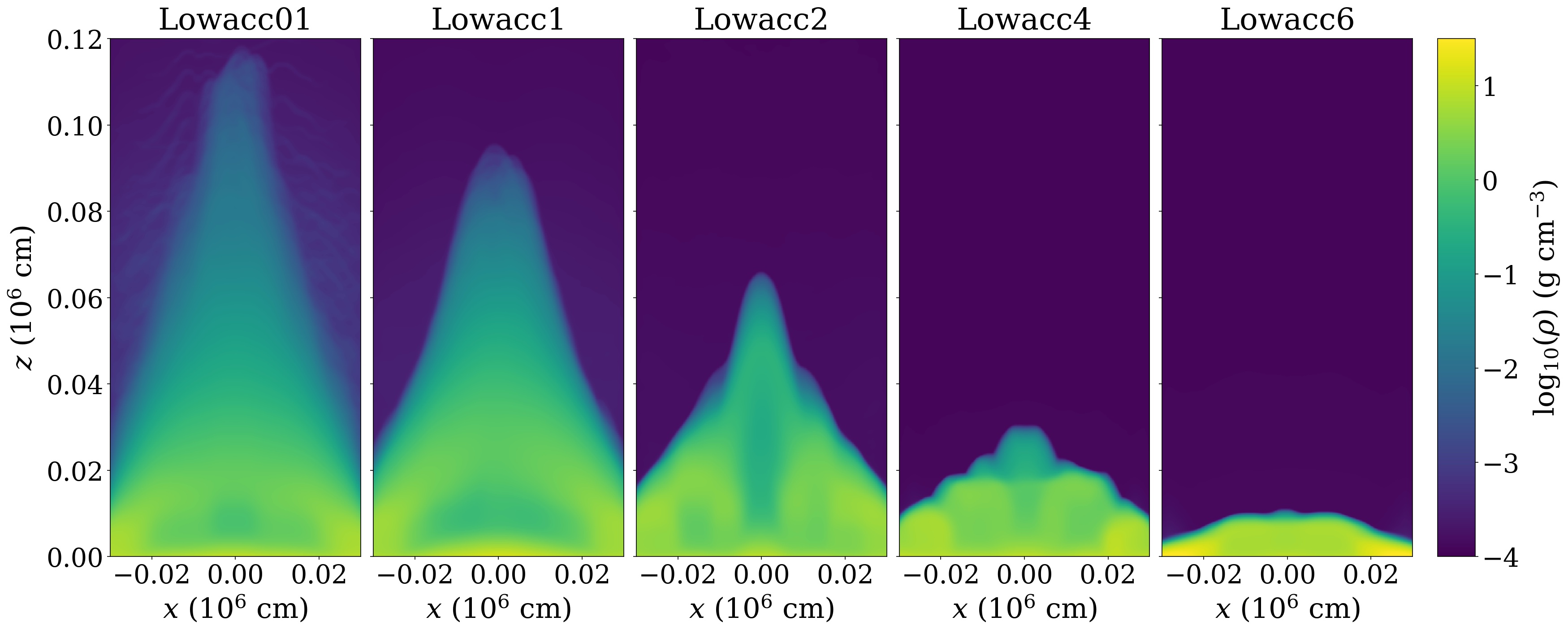}
    \caption{Time-averaged density profiles of simulations Lowacc01 to Lowacc6 from $t=6000t_{\rm sim}$ to $t=13000t_{\rm sim}$.  The accretion rate is the same in all these simulations, and the only thing that changes is the magnetic field used to determine the opacity, increasing from $10^{11}$~G on the left to $6\times10^{12}$~G on the right.}
    \label{fig:average_density}
\end{figure*}

\begin{figure*}
    \centering
    \includegraphics[width=0.8\linewidth]{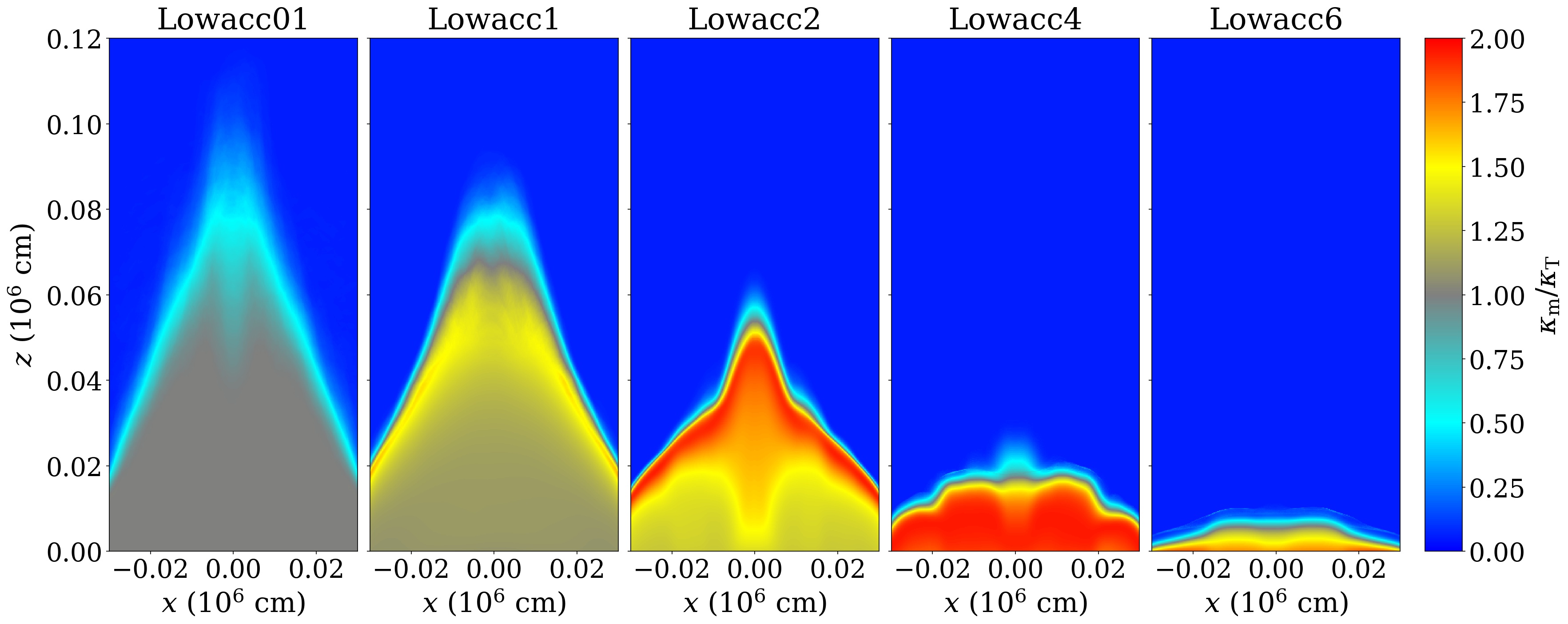}
    \caption{Time-averaged opacity profiles of simulations Lowacc01 to Lowacc6 from $t=6000t_{\rm sim}$ to $t=13000t_{\rm sim}$.
    }
    \label{fig:average_opacity}
\end{figure*}

In this section, we use the five low accretion rate simulations (Versions~1 to 5, Lowacc01 - Lowacc6) to study how the dynamical behavior of the accretion column varies with changes in opacity caused by changes in magnetic field strength, at fixed accretion rate.  Naively, one might expect that increasing the magnetic field would tend to decrease the overall opacity, so the radiation pressure force on the plasma would decrease, and the time-averaged height of the accretion column would therefore decrease.  The time-averaged density profiles for the low accretion rate simulations are shown in \autoref{fig:average_density}, and do in fact show a decrease in column height with increasing field strength.  However, for this accretion rate, \autoref{fig:average_opacity} shows that the opacity actually generally increases, not decreases, in the time-averaged column structure as the magnetic field increases from Lowacc01 to Lowacc4.  Only in going from Lowacc4 to Lowacc6 does the opacity
decrease, and the column then almost becomes a surface hot spot.

Once again, \autoref{fig:opacity} provides the explanation for this behavior.  Although the shapes of the opacity curves are all very similar in this logarithmic plot, the actual linear temperature width is very small at low magnetic field strengths and increases toward higher field.  For the lowest magnetic field (Lowacc01 in the left-most panel of \autoref{fig:average_opacity}), the opacity across the shock jumps right through the peak into the Thomson regime.  The low values of the opacity
at the surface, indicated by the cyan color in \autoref{fig:average_opacity}, are simply due to the time-averaging over the vertical oscillation of the shock, so that at those locations one is averaging over the near-zero opacity in the free-fall zone and the Thomson opacity in the sinking zone.  As we discussed in the previous section, Lowacc1 transitions through the opacity peak to almost reach Thomson in the interior, but the time-averaged opacity structure (second panel from the left in \autoref{fig:average_opacity}) does not exhibit the actual peak opacity because of the averaging over lower opacities in the vertically oscillating shock.  As one moves to the simulations with still higher magnetic field, the wider and wider opacity peak becomes better and better resolved in the time-average, and the opacity in the deep interior never declines back down to Thomson. In Lowacc4, the opacity is close to the peak value everywhere in the sinking zone, and finally in Lowacc6 the temperature jump across the shock and down to the base of the column is too small for the opacity to quite reach the peak.  (Note the location of the intersection of the base temperature indicated by the left vertical line in \autoref{fig:opacity} with the $6\times10^{12}$~G magnetic opacity curve.)  The opacity therefore increases as one moves downward toward higher temperatures in what is left of the sinking zone.  Further increase of the magnetic field would evidently result in too little post-shock opacity to support a column, and we would be left with a hot spot.

We are still left with the question as to why the column height decreases with increasing magnetic field in going from Lowacc1 to Lowacc4, even while the opacity is increasing inside the column.  Let us begin by comparing Lowacc1 and Lowacc2.  Lowacc2 has significantly increased opacity just below the shock compared to Lowacc1, simply because it has a wider opacity peak.  This appears to provide a further barrier to vertically distributing the accretion power liberated in the shock to the rest of the column.  Instead, more of the accretion power is radiated outward from the shock. This results in a shorter column, and a column that does not oscillate vertically as much as in the weaker magnetic field case.  As we discussed above, these vertical oscillations are also a direct way of redistributing the accretion power vertically, but this also is now failing in the shorter column.  As we continue to increase the magnetic field from Lowacc2 to Lowacc4 the post-shock opacity is even larger, and the column height again decreases, becoming almost a hot spot configuration.   A concomitant feature of our simulations is that the vertical displacement amplitude of the oscillation declines with increasing magnetic field strength at fixed accretion rate.  As we discuss in more detail in \autoref{subsec:Variability} below, this, in turn, results in a smaller luminosity variability amplitude.

The postshock opacity declines in moving from Lowacc4 to Lowacc6, because the opacity peak is now so wide that the peak is not reached in Lowacc6.  And yet the column height still decreases.  This is due to a second important contribution to the decrease in column height which is evident from \autoref{tab:results}.  Despite the fact that simulations Lowacc01-Lowacc6 have the same accretion rate, they do not have the same emitted luminosity.  In fact, the luminosities of the simulated columns show a decreasing trend with stronger magnetic field, except in going from Lowacc1 to Lowacc2, where the luminosity slightly increases.  The reason for this is that as the column height declines (driven by the postshock opacity variation), the sideways emitting area declines and, at the same time, the shock-liberated accretion power is brought closer to the stellar surface.  This means that more of the accretion power is able to advect through the sinking zone and into the relatively cold neutron star base layer, and this is also a contributing factor to the decrease of the column height.  It is this effect that dominates the decrease in going from Lowacc4 to Lowacc6.  In going from Lowacc1 to Lowacc2, the post-shock opacity reaches its maximum possible value, and this is what causes Lowacc2 to have a slightly higher fraction of radiated accretion power.

\subsection{Effects Arising from Changing Accretion Rate}
\label{subsec:Effects_from_the_Accretion_Rate}

\begin{figure*}
    \centering
    \includegraphics[width=0.8\linewidth]{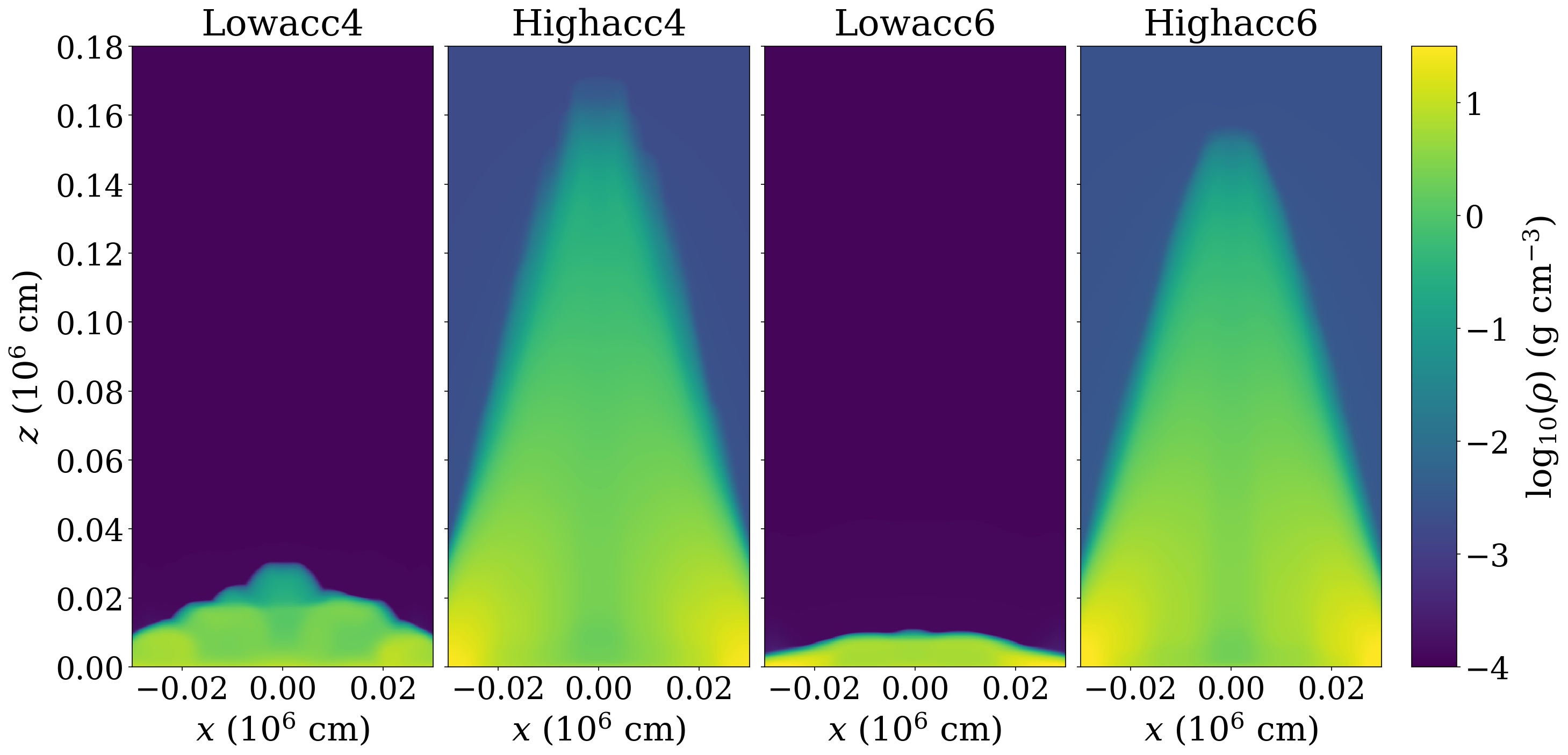}
    \caption{Comparison of the time-averaged density profiles of simulations with the same magnetic field strength but different accretion rates, from $t=6000t_{\rm sim}$ to $t=13000t_{\rm sim}$.  The two panels at the left are for $4\times10^{12}$~G magnetic fields, and the two on the right are for $6\times10^{12}$~G.}
    \label{fig:average_density_with_different_rate}
\end{figure*}

\begin{figure*}
    \centering
    \includegraphics[width=0.8\linewidth]{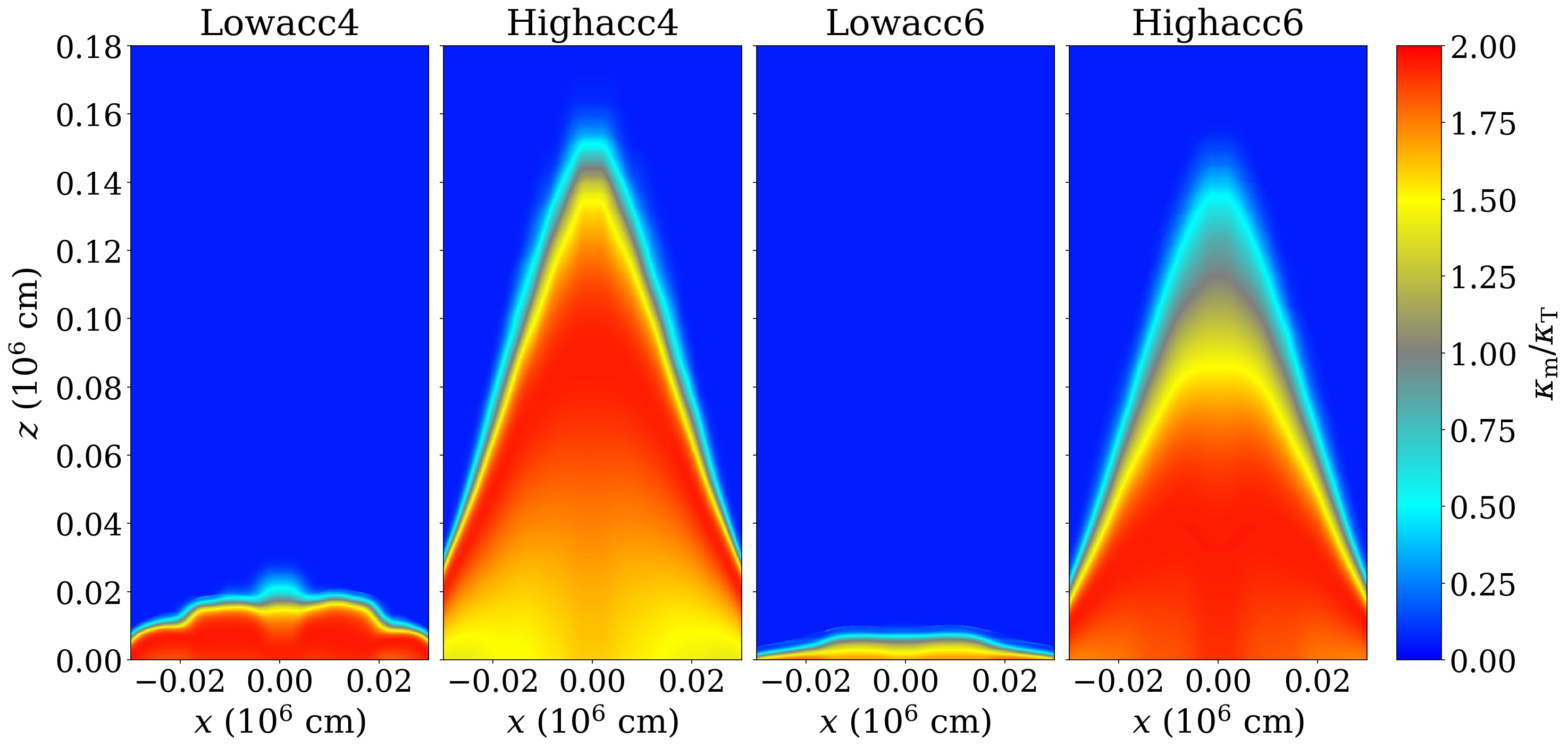}
    \caption{Time-averaged magnetic opacity profiles of simulations with the same magnetic field strength but different accretion rates, from $t=6000t_{\rm sim}$ to $t=13000t_{\rm sim}$}
    \label{fig:average_opacity_with_different_rate}
\end{figure*}

\begin{figure*}
    \centering
    \includegraphics[width=0.8\linewidth]{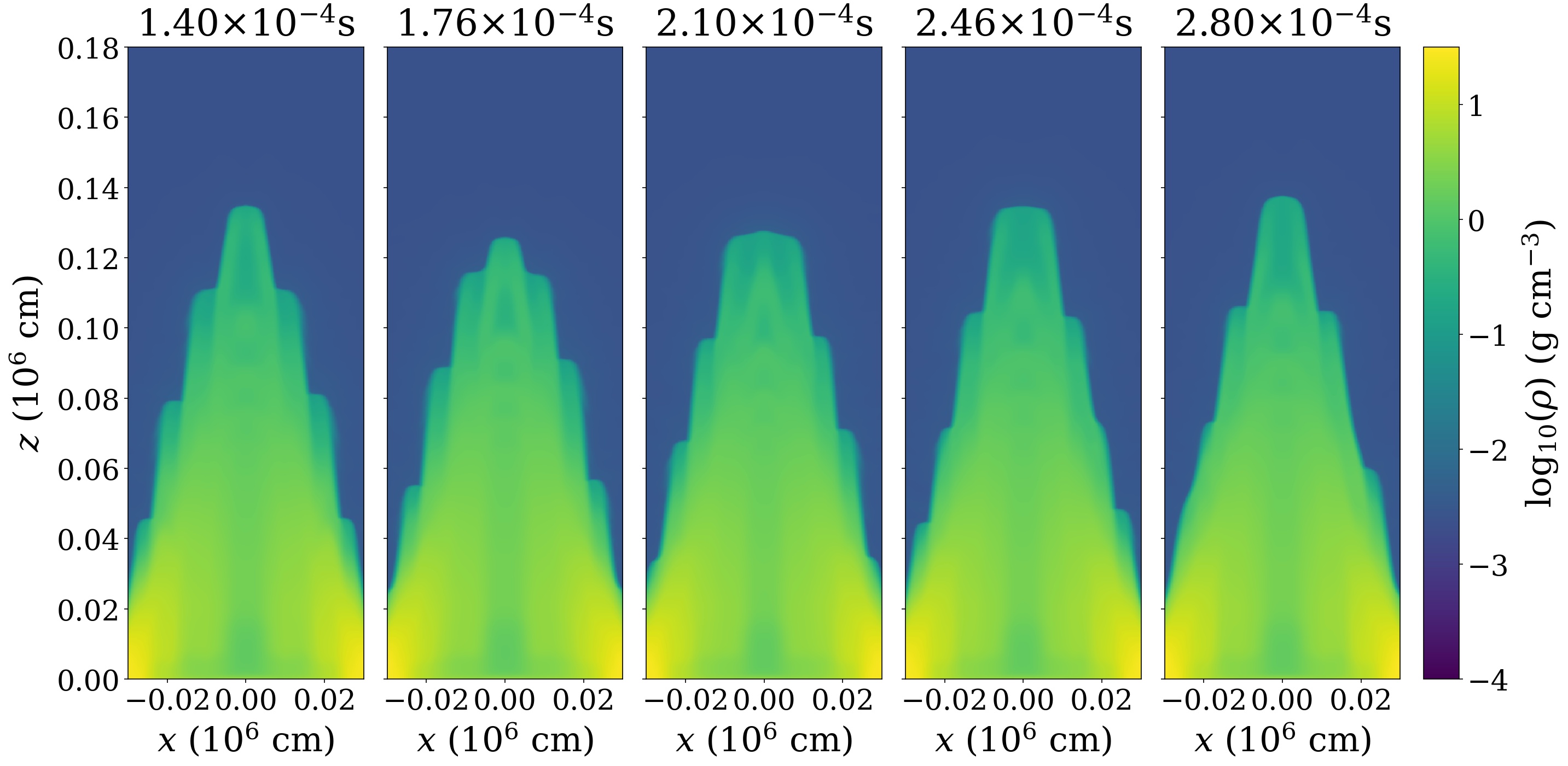}
    \includegraphics[width=0.8\linewidth]{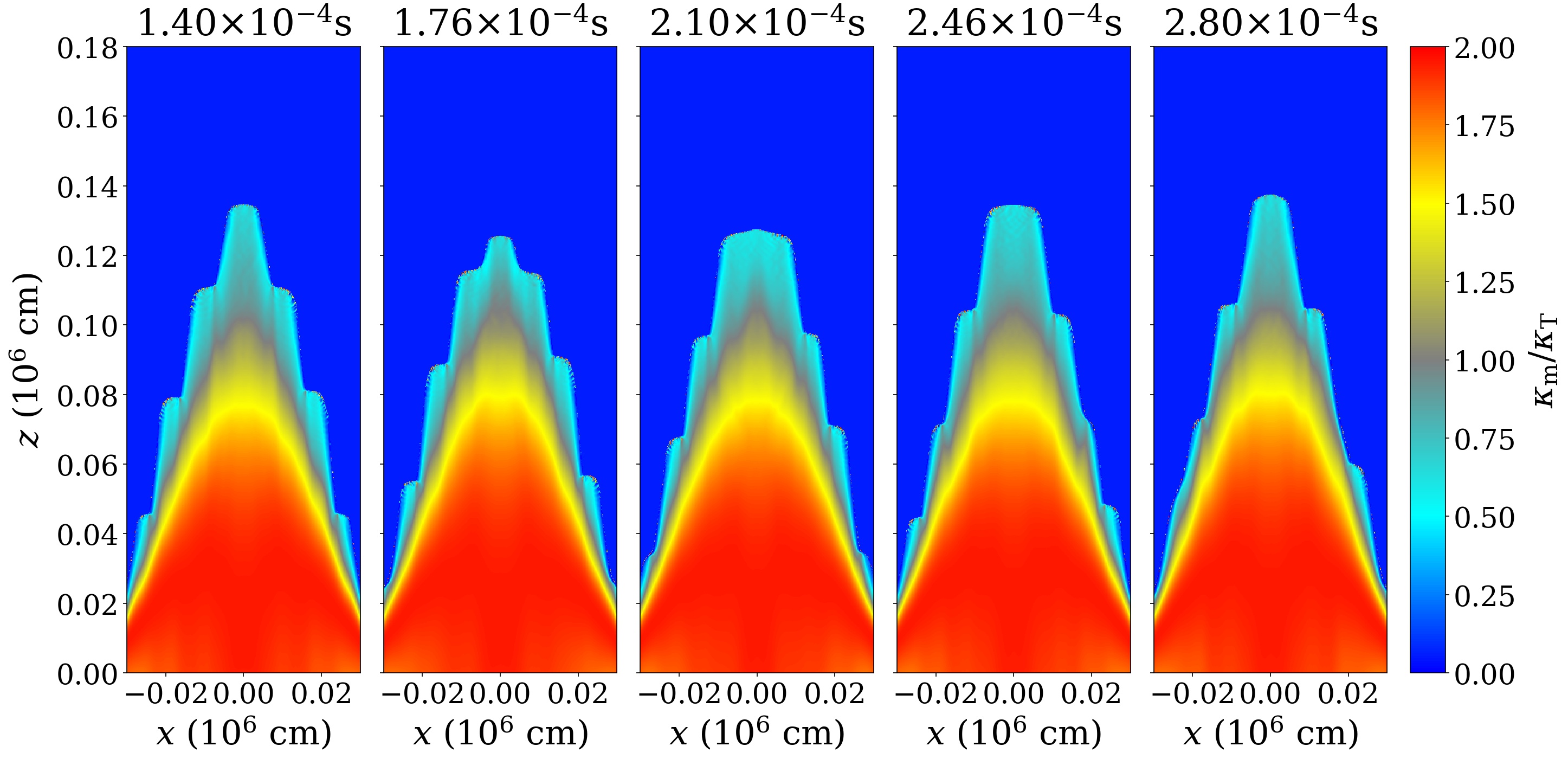}
    \caption{Snapshots of the density distribution and opacity distribution in simulation Highacc6 over an interval of $1.4\times10^{-4}$~s.  This corresponds to approximately one oscillation period in the light curve for this simulation.
    }
    \label{fig:sinking_region_oscillationh06}
\end{figure*}

\begin{figure*}
    \centering
    \includegraphics[width=0.8\linewidth]{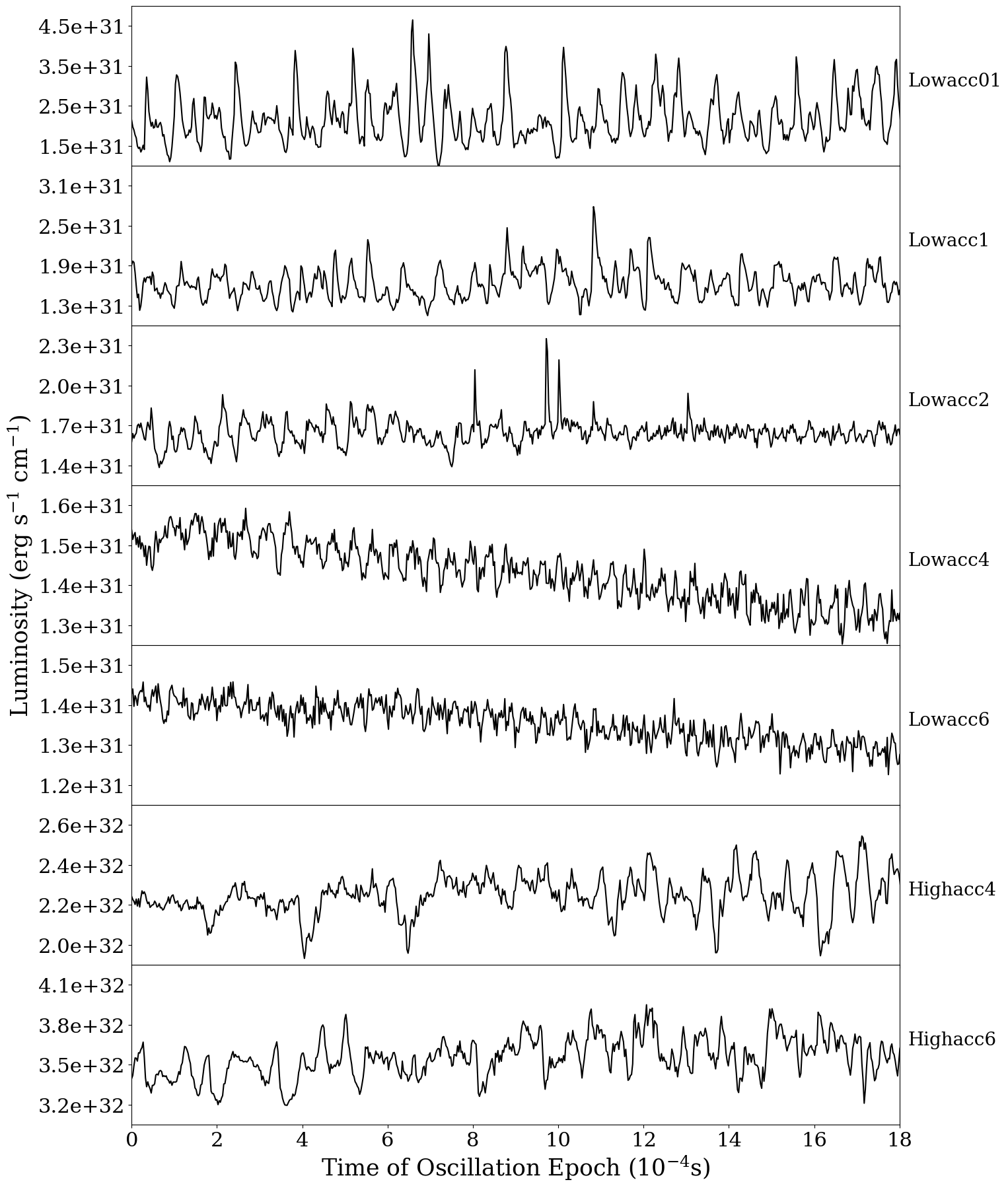}
    \caption{Light curve variation of all simulations, starting from times when the simulations achieved an approximate steady state with persistent oscillations (6000~$t_{\rm sim}$ for the low accretion rate simulations Lowacc01-Lowacc6, 11000~$t_{\rm sim}$ for the Highacc4 and Highacc6 simulations.}
    \label{fig:light_curve}
\end{figure*}

Our highest field strength simulations at low accretion rate (Lowacc4 and Lowacc6) resulted in very short accretion columns.  One would expect that increasing the accretion rate in these two magnetic field regimes would produce more luminosity and radiation pressure support, resulting in taller columns.  We did this in simulations Highacc4 and Highacc6, and we present the resulting time-averaged density and opacity in
\autoref{fig:average_density_with_different_rate} and \autoref{fig:average_opacity_with_different_rate}, respectively.  Highacc4 has the same magnetic field ($4\times10^{12}$~G) as Lowacc4, but an accretion rate that is 15 times larger.  Highacc6 has the same magnetic field ($6\times10^{12}$~G) as Lowacc6, but an accretion rate that is 20 times larger. As expected, the increased accretion rates in these two simulations result in taller structures, and in fact taller than the weak magnetic field, low accretion rate simulation Lowacc01.  As indicated in \autoref{tab:results}, no more than ten percent of the accretion power is transferred to the neutron star base, compared to $\simeq40$~percent in the low accretion rate simulations Lowacc4 and Lowacc6.  A taller column with its much larger surface area (and higher postshock opacity in these two cases) is better able to release the accretion power in emergent radiation.

\autoref{fig:sinking_region_oscillationh06} shows the time-dependent behavior of both the density and the opacity over a time interval of $1.4\times10^{-4}$~s in simulation Highacc6.  As we discuss in more detail in the next subsection, this corresponds to one oscillation period in the peak frequency of the power spectrum of the light curve of this simulation.  Comparing \autoref{fig:sinking_region_oscillationh06} to \autoref{fig:sinking_region_oscillation01} and \autoref{fig:sinking_region_oscillation1}, it is apparent that the structure of the column is varying much more dramatically in the latter, weaker magnetic field simulation.  This is true even though these simulations produce columns with comparable time-averaged heights.  We discuss why this is in the next subsection.

\subsection{Variability}
\label{subsec:Variability}

\autoref{fig:light_curve} depicts the luminosity light curves of all the simulations.  We compute these light curves by simply summing the horizontal and vertical lab-frame fluxes times cell face areas for cells near the photosphere.  Because the velocity of the flow is restricted by the magnetic field to be almost exactly vertical, there is almost no difference between the lab frame and fluid frame horizontal radiation fluxes.  However, for simulations Highacc4 and Highacc6, which ran for a much longer time than the Lowacc simulations, some horizontal motion began to occur as the effective boundary at the base overheats and we start to lose magnetic confinement.  This produces artificial high frequency, Alfv\'enic oscillations in the lab frame flux and we have removed these from the light curves in \autoref{fig:light_curve} by using the fluid frame horizontal flux.

\begin{figure}
    \centering
    \includegraphics[width=1.0\linewidth]{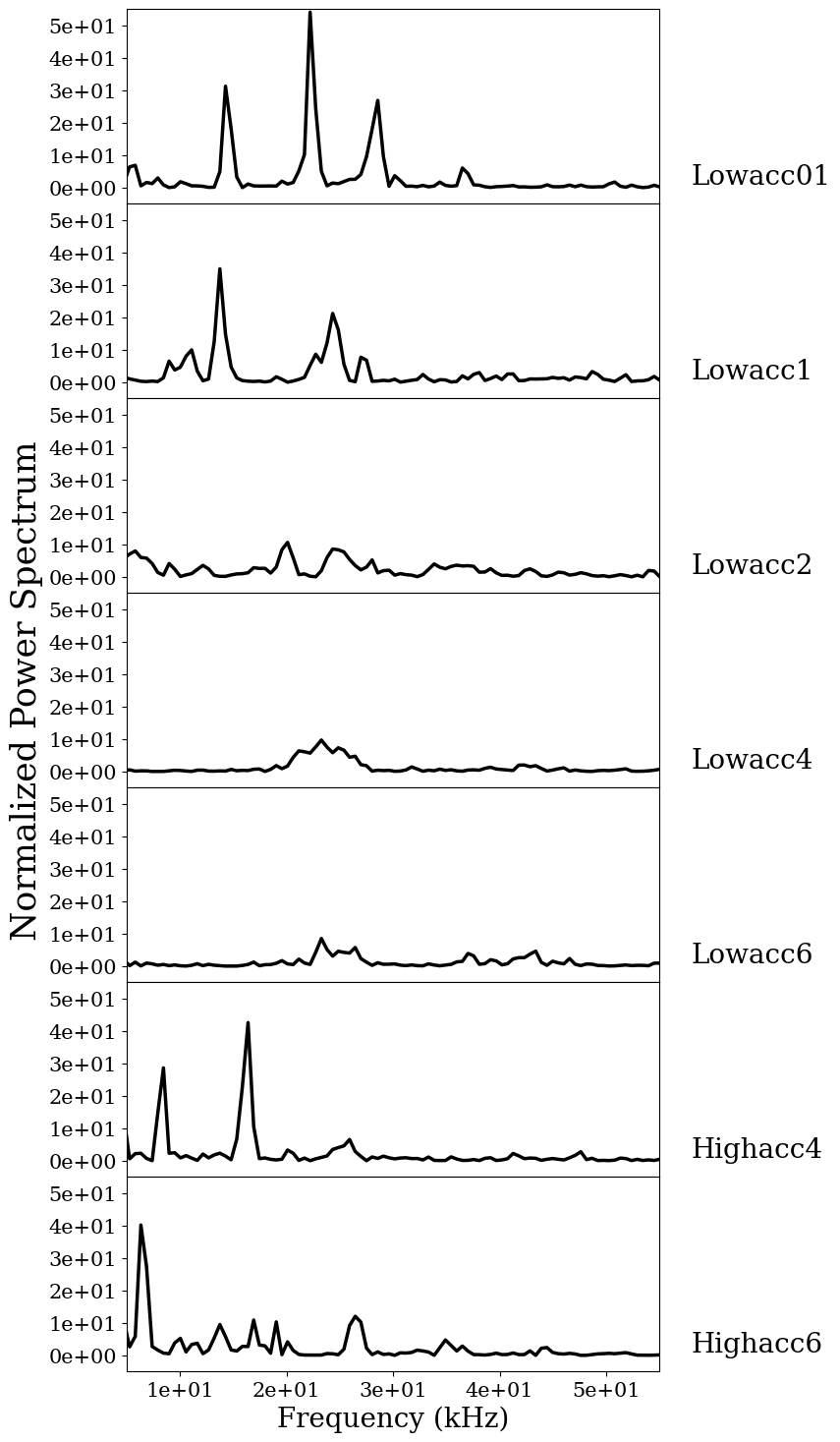}
    \caption{Power spectrum of each simulation computed using Welch's method. 
    }
    \label{fig:power_spectrum}
\end{figure}

\begin{figure}
    \centering
    \includegraphics[width=1.0\linewidth]{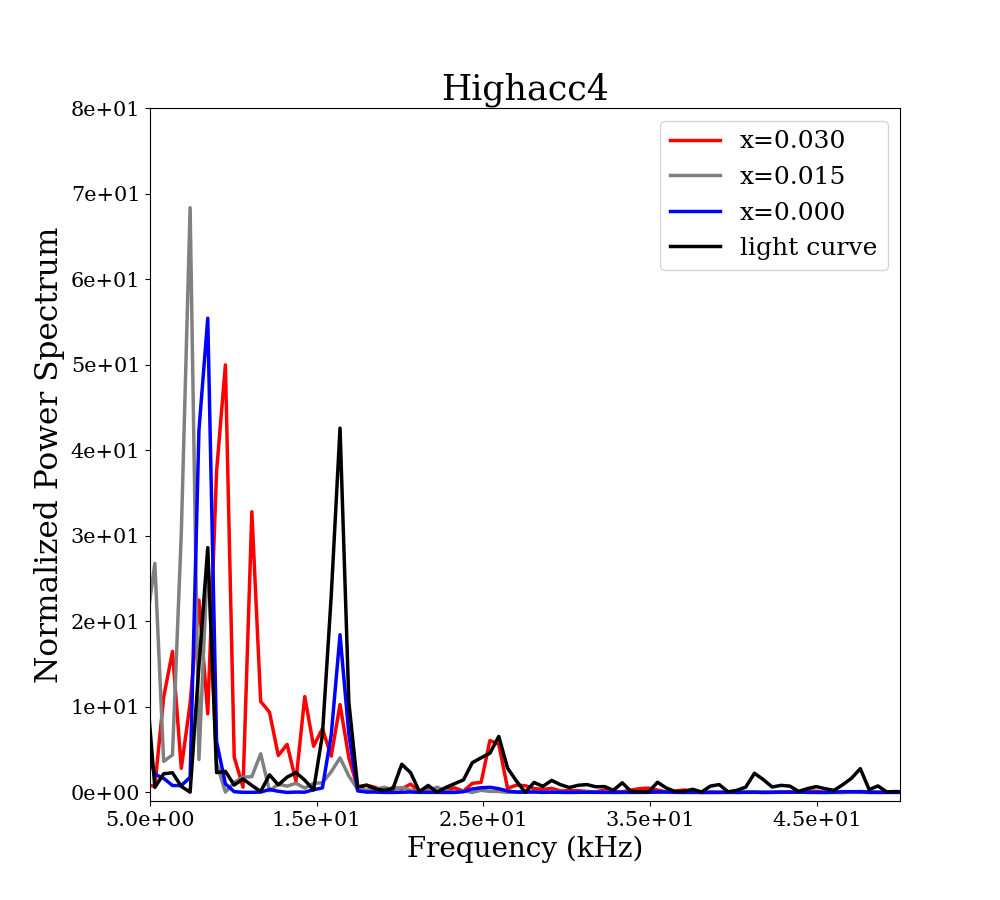}
    \caption{Power spectrum of shock height as a function of horizontal position for simulation Highacc4.}
    \label{fig:power_spectrum_shock_height}
\end{figure}

All the light curves in \autoref{fig:light_curve} show significant high frequency variability, including quasi-periodic oscillations (QPOs) with varying degrees of coherence.  Power spectra of these light curves are shown in \autoref{fig:power_spectrum}, and the relative amplitude of the most significant QPOs are listed in \autoref{tab:oscillation}.  In our previous Thomson scattering simulations \citep{Paper2, Paper3}, we showed that the origin of these oscillations is due to a breakdown in thermal equilibrium caused by the fact that advection of heat in the settling flow and $PdV$ work are generally unable to balance the sideways radiative cooling when the column is at maximal vertical extent.  The column therefore overcools and the shock height falls, resulting then in overheating which causes the shock to rise again.  This can vary with horizontal position within the column as separate vertical fingers oscillate up and down.  That this is happening here in these magnetic opacity simulations is shown in \autoref{fig:power_spectrum_shock_height}, which compares the power spectrum of shock height variations at different horizontal locations in Highacc4 with the luminosity power spectrum.  Clearly the three most significant QPOs in the latter match the shock height oscillations at different horizontal locations.  Note that more light curve power is in the middle frequency QPO, while the most shock height power is in the lowest frequency QPO.  These two QPOs are in a $2:1$ frequency ratio, and the relationship between shock height and emitted light is not likely to have the same proportionality at different harmonics in these non-sinusoidal oscillations.

\begin{table}
	\centering
	\caption{Time-averaged radiated luminosity as fraction of accretion power, standard deviation of light curve variability, and two methods of to determine the fan beam fraction:  the ratio of mean horizontal luminosity to mean luminosity, and mean of the horizontal luminosity fraction. }
	\label{tab:oscillation}
	\begin{tabular}{clcccc} % four columns, alignment for each
		\hline
		Name & $\left<L\right>/L{\mathrm{acc}}$ & $\sigma(L)/\left<L\right>$ & $\left<L_x\right>/\left<L\right>$ & $\left<L_x/L\right>$\\
		\hline
		Lowacc01 & 0.875 & 0.271 & 0.843 & 0.858 \\
		Lowacc1  & 0.659 & 0.139 & 0.890 & 0.892 \\
		Lowacc2  & 0.666 & 0.056 & 0.966 & 0.965 \\
        Lowacc4  & 0.577 & 0.054 & 0.375 & 0.373 \\
        Lowacc6  & 0.548 & 0.036 & 0.138 & 0.137 \\
        Highacc4 & 0.909 & 0.044 & 0.956 & 0.956 \\
        Highacc6 & 0.961 & 0.042 & 0.986 & 0.987 \\
		\hline
	\end{tabular}
	\label{tab:results}
\end{table}

Consistent with their physical origin, the frequencies of these oscillations are related (inversely) with the local cooling time at that horizontal section of the column.  For example, for taller accretion columns, the shock front generally needs to oscillate with a sufficiently large vertical amplitude to replenish the heat to support the bottom region, and this also leads to a longer oscillation time and lower oscillation frequency.  This is exactly what we see in \autoref{tab:results} and \autoref{fig:power_spectrum} as the stronger magnetic field in simulations Lowacc01 to Lowacc6 results in shorter accretion columns.   In particular, when the magnetic field is sufficiently strong (e.g. Lowacc4 and Lowacc6), the accretion column almost collapses into a hot spot and then oscillates fast with a small amplitude.  Since the sideways cooling area is small, the variations in the light curve are very small. However, when we restore taller column heights by increasing the accretion rates in these high magnetic field cases (Highacc4 and Highacc6), oscillations with lower frequencies occur.

However, the oscillation amplitude in the high accretion rate simulations is substantially less than that in the low accretion rate simulations of comparable height, particularly Lowacc01.  This is true both in luminosity, and in the overall shock height variation.  In fact, the main body of the sinking zone away from the oscillating shock has an almost static structure apart from the presence of horizontally-propagating entropy waves.  The high accretion rate simulations have higher horizontal optical depth (by one to two orders of magnitude throughout most of the column in the case of Highacc6) than in Lowacc01, both because the density is higher and because the opacity is larger.  Hence the cooling time is longer and the column is better able to establish a balance between heating from accretion power and radiative cooling.  It is only in the upper part of the column where thermal equilibrium is unable to be established because of the more rapid diffusive cooling, and this is the region that oscillates. While the increased opacity in the lower, high temperature regions enables these regions to achieve thermal balance, it turns out that this temperature dependence also leads to a new unstable behavior at late times, and we discuss this further below in \autoref{subsec:opacity_instability}.

\subsection{Angular Distribution of Emergent Radiation}
\label{subsec:Angular_Distribution_of_Emergent Radiation}

\begin{figure}
    \centering
    \includegraphics[width=\linewidth]{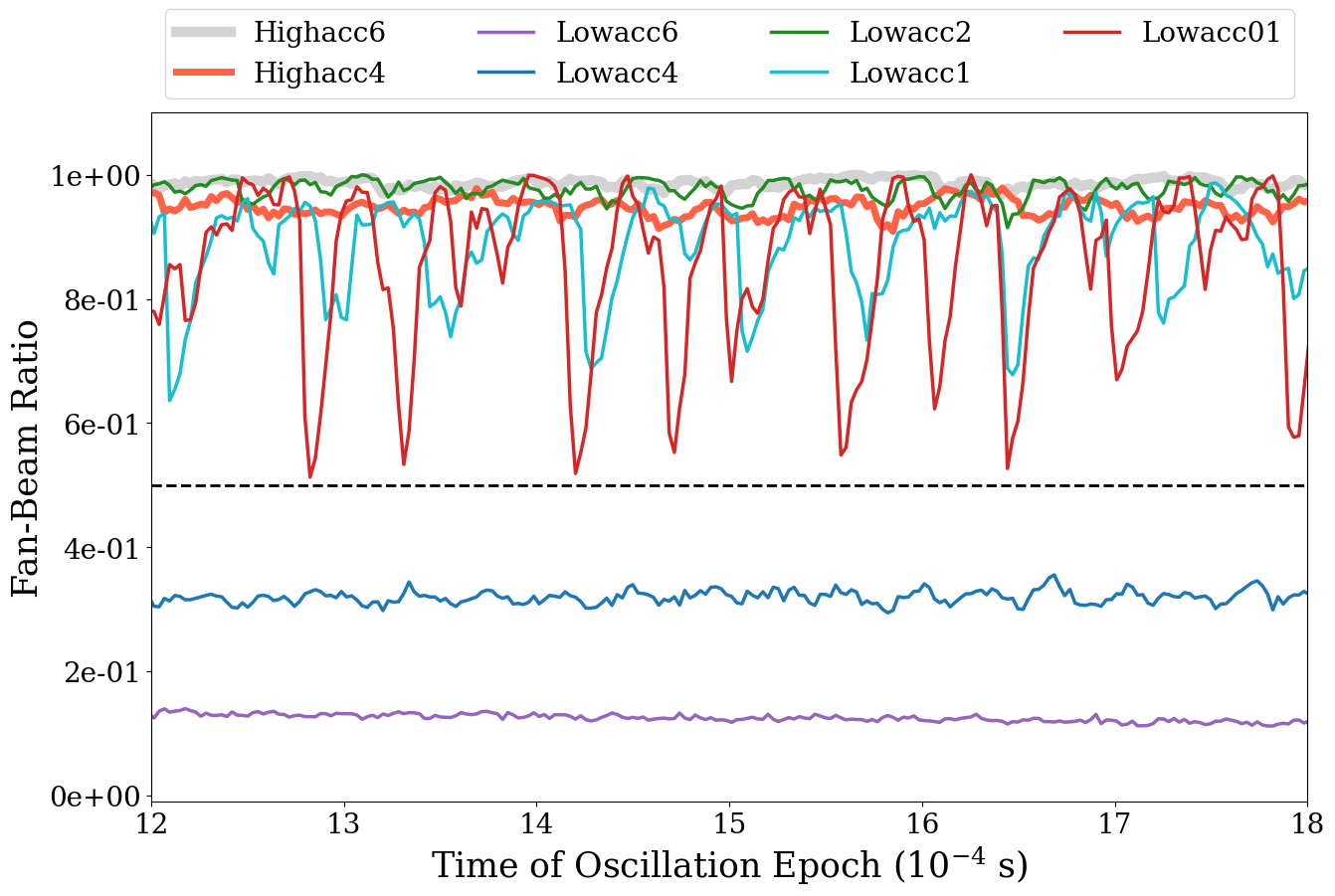}
    \caption{Ratio of the side luminosity to the total luminosity as a function of time in all seven simulations.  The time here corresponds to epochs 12-18$\times10^{-4}$~s in \autoref{fig:light_curve}.}
    \label{fig:light_curve_ratio}
\end{figure}

When the neutron star accretion forms a hot spot at a relatively low accretion rate or in a strong surface magnetic field, the accreting material is halted at the stellar surface and releases mechanical energy into radiation.  Since the incoming flow is cold and has low opacity, it is transparent and radiation can directly leave the system upwards (i.e. pencil beam).  However, when the accretion rate is sufficiently high, the accretion flow is shocked above the stellar surface and forms a radiation pressure supported, optically thick region below which radiation emerges from the sides  (i.e. fan beam).

The fact that classical models of accretion columns produce fan beam radiation patterns is simply due to the fact that most of the emission area is on the sides, even though the accretion shock itself is at the top of the column.  In our 2D, more mound-shaped columns, the sides still have more emitting area, and the shock itself covers this mound-shape, so that significant direct dissipation of the accretion power is also happening along the sides of the column.

The angular distribution of the emitted radiation is therefore of course determined by both the geometry and surface brightness distribution of the photosphere of the column.  In \autoref{fig:light_curve_ratio}, we compare the instananeous fraction of emergent radiation that is in a fan-beam across all the simulations, computed from the integral of horizontal flux leaving the photosphere divided by total luminosity.  We also list the time-averaged fan beam ratio in \autoref{tab:results}, computed in two different ways: the ratio of time-averaged horizontal luminosity to time-averaged total luminosity, and the time-average of the ratio.  Despite the strong variability exhibited in \autoref{fig:light_curve_ratio}, particularly in Lowacc01 and Lowacc1, these two methods of averaging produce very consistent results.

For simulations Lowacc4 and Lowacc6 with short, flat columns (almost hot-spots), the fan-beam fraction is significantly below 0.5 (e.g. Lowacc4 and Lowacc6) and the emergent radiation is therefore more like the classical pencil-beam of 1D models. In all the other simulations, the system develops a columnar structure, where most radiation escapes sideways in a fan-beam pattern (i.e. fan-beam fraction above 0.5).  However, the low-accretion columns in relatively low magnetic fields (Lowacc01 and Lowacc1) exhibit large variations of radiation beaming patterns, which result from the large shock oscillation amplitude.  In particular, when the accretion column is mostly compressed (e.g. middle panel in \autoref{fig:sinking_region_oscillation01}), the sideways cooling is minimized and the system is over-heated with quite a large fraction of radiation leaving from the top.  For similar heights of columnar structure, the variation of the fan-beam fraction in the case of high accretion rates and strong magnetic fields (Highacc4 and Highacc6) is in general smaller because the shock front oscillates with much smaller amplitude.

\subsection{Opacity-Driven Instability in the High Accretion Rate Simulations}
\label{subsec:opacity_instability}

\begin{figure*}
    \centering
    \includegraphics[width=1.0\linewidth]{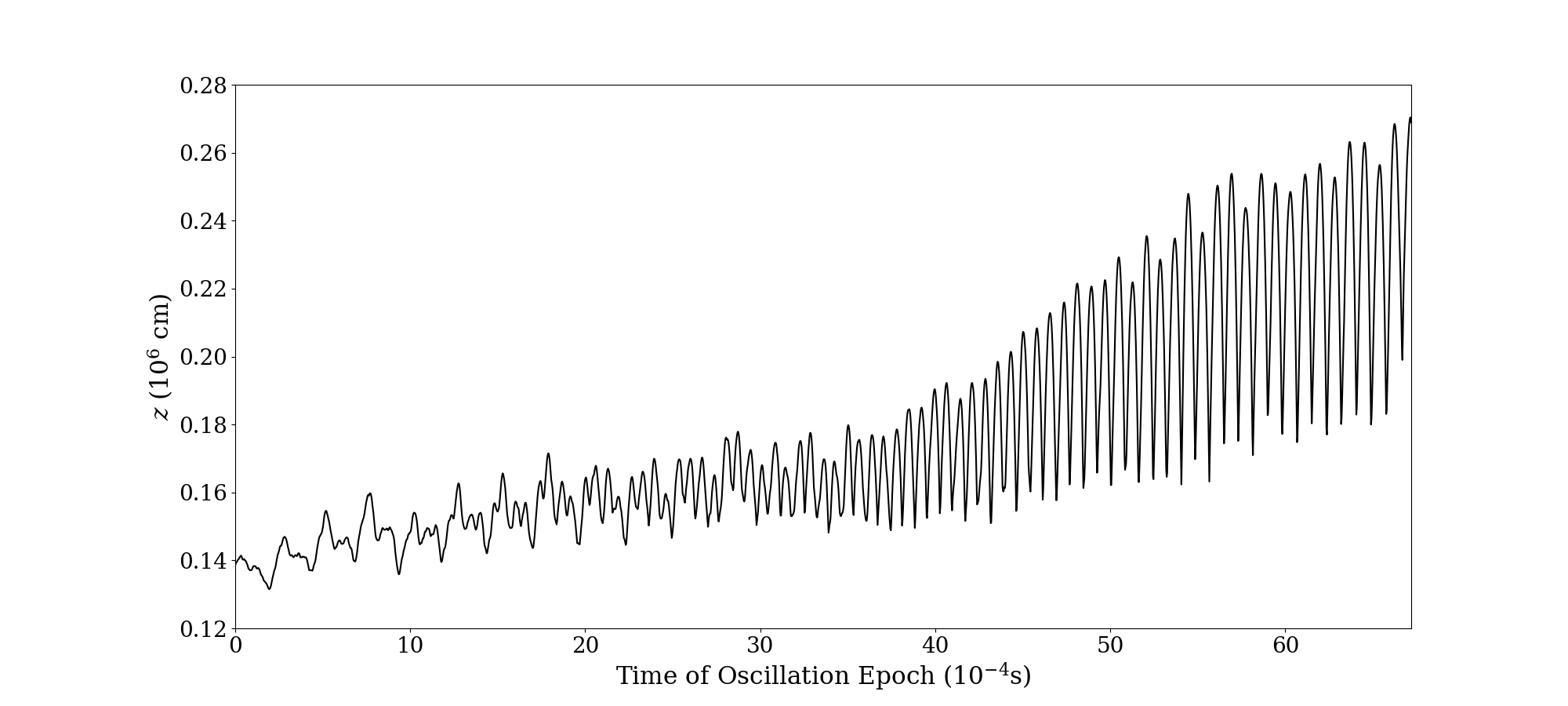}
    \caption{Height of the accretion shock at the $x=0$ middle of the column as a function of time for simulation Highacc4.  The time axis is the same as in \autoref{fig:light_curve}, but extended for much longer.
    }
    \label{fig:late_shock_osc_highacc4}
\end{figure*}

\begin{figure}
    \centering
    \includegraphics[width=1.0\linewidth]{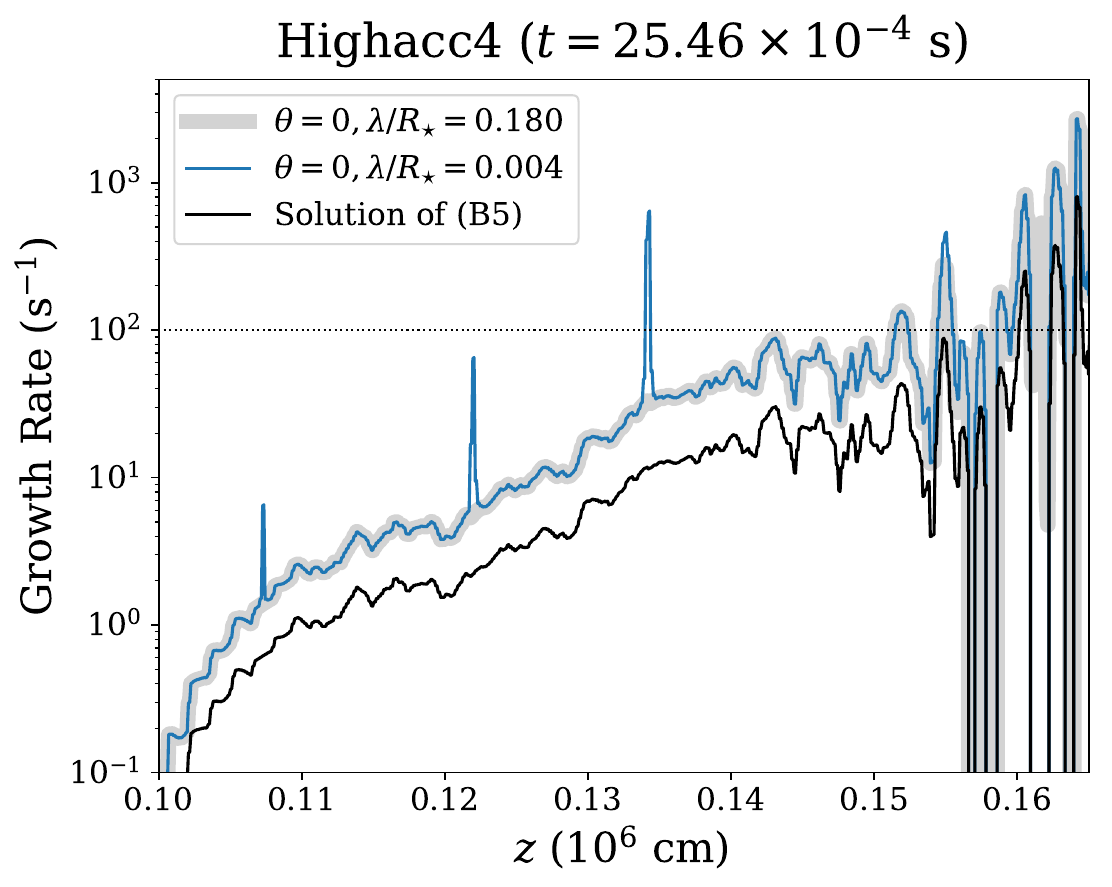}
    \caption{Predicted growth rate of the instability driven by the temperature-dependent magnetic opacity from linear instability theory in a static medium
    (\hyperref[appendix:pbi_mag_opacity]{Appendix B}).  The solution is consistent with the rough growth time scale of the oscillation amplitude that is evident in \autoref{fig:late_shock_osc_highacc4}, indicated by the horizontal dotted line. 
    }
    \label{fig:mag_instability_growth}
\end{figure}

\autoref{fig:late_shock_osc_highacc4} shows the temporal behavior of the shock height at the middle of the column for simulation Highacc4, extending well beyond the time range shown in \autoref{fig:light_curve}.  Both the shock oscillation amplitude and the average height of the shock increase dramatically beyond $t=40\times10^{-4}$~s.  Simulation Highacc6 also shows evidence of this behavior, though we were unable to track it for as long as Highacc4.  We never observed such behavior in any of our pure Thomson scattering simulations \citep{Paper2,Paper3}.  This suggests the presence of an additional unstable mechanism in the column dynamics that is directly related to the temperature-dependence of the magnetic scattering opacity.

Under conditions of pure Thomson scattering and a vertical magnetic field, the maximum growth rate of the photon bubble instability occurs for near horizontal propagation directions in the slow diffusion regime \citep{Paper1} and propagation directions at 45 degrees in the rapid diffusion regime \citep{gam98}.  As we show in \hyperref[appendix:pbi_mag_opacity]{Appendix B}, a magnetic opacity that increases with temperature modifies the photon bubble dispersion relation such that unstable growth can occur for vertical propagation in the slow diffusion regime.  The physics of this unstable growth differs from the pure photon bubble instability, and while it is slower than the photon bubble instability, we suggest that it is the cause of the growth that takes place at late times in \autoref{fig:late_shock_osc_highacc4}. 

Because of the broad temperature width of the magnetic opacity peak in both Highacc4 and Highacc6, the postshock material does not reach the opacity peak and the opacity therefore increases further with temperature and depth in the column (see \autoref{fig:average_opacity_with_different_rate} and \autoref{fig:sinking_region_oscillationh06}).  The time-averaged opacity structure in simulation Lowacc1 shown in \autoref{fig:average_opacity} also appears to show a region of opacity increasing with temperature, but this is an artifact of the time-averaging.  The instantaneous opacity structure shown in \autoref{fig:sinking_region_oscillation1} shows no such behavior.  Hence only Highacc4 and Highacc6 have extended regions in which the opacity is below the peak and therefore increases with temperature, and it is only these simulations that exhibit this unstable behavior.

We have solved the full dispersion relation \autoref{eq:thetadisp} for conditions along the $x=0$ midline for a snapshot of Highacc4 at $t=25.46\times10^{-4}$~s, shortly before the unstable behavior in \autoref{fig:late_shock_osc_highacc4} becomes evident. The results are shown in \autoref{fig:mag_instability_growth} as a function of height in the column, for two different wavelengths and for vertical wave vectors (angle between wave vector and magnetic field $\theta=0$). We also solved the approximate dispersion relation \autoref{eq:mag_instability_approx} for this instability, which is wavelength-independent, and the result is very similar. The unstable region in the center of the column covers exactly the range of heights where the opacity increases with depth and temperature.  Growth rates near the top of the column are $\sim100$~s$^{-1}$, and this is consistent with the growth time scale $\sim10^{-2}$~s in \autoref{fig:late_shock_osc_highacc4}. 

All of this is based on short wavelength WKB theory and so is difficult to apply to our actual column structure.  Moreover, the central shock of the accretion column is already highly dynamic due to the basic thermal oscillation driven by the mismatch between heating and cooling, so linear perturbation theory on a static structure cannot be applied.  Nevertheless, the predicted growth rates are comparable to what we see in these simulations, and it is suggestive that only Highacc4 and Highacc6 exhibit this behavior, and only Highacc4 and Highacc6 have vertically extended regions at lower temperatures than where the peak opacity occurs.  It is important to note that our opacities depend only on temperature, and not density.  More accurate opacities \citep{Suleimanov2022} will have some density dependence, and these can also excite instability, even in the rapid diffusion regime (\hyperref[appendix:pbi_mag_opacity]{Appendix B}).

\section{Discussion}
\label{sec:discussion}

\subsection{Numerical Caveats}

We remind the reader that we treat the neutron star as a classical gas pressure dominated region and only allow heat transport by advection and radiation transport \citep{Paper2}.  This effective boundary condition is designed to minimize the boundary effects on the overall column dynamics.  However, the fact that advection of heat into the neutron star is a contributing factor to the overall height of the shorter column simulations Lowacc4 and Lowacc6 suggests that this boundary condition might affect the overall scaling between accretion rate and column height. Our treatment of the neutron star surface neglects degeneracy pressure, which affects the heat capacity, and also neglects thermal conduction by electrons.  It would be worthwhile in future to incorporate a more accurate treatment of the interface between the accretion column and the neutron star when the columns are approaching the hot spot regime, in order to more accurately determine how the column height varies with accretion rate.  However, we suspect that the overall dynamics of the column will not be affected significantly.

At very late times, both Highacc4 and Highacc6 exhibit sudden flares of radiation from one side of the base of the column, and this rapidly causes the simulation to crash.  This is an unphysical behavior that arises from the effective lower boundary condition, where the neutron star surface has heated sufficiently that radiation pressure starts to bend the magnetic field.  As we mentioned in \autoref{subsec:additional_numerical_treatments}, the field we use in the MHD is lower than the field used to determine the opacities in order to avoid errors in the conservative to primitive variable inversion.  Further work is needed to implement a variable inversion that would allow us to run at higher magnetic fields, in addition to improving the physics of the bottom boundary condition.  This would better enable us to determine the long-term outcome of the opacity driven instability that manifested in these simulations at late times.

\subsection{Opacity Due to Pair Production}
\label{subsection:pair_production}

\begin{figure}
\centering
\includegraphics[width=\linewidth]{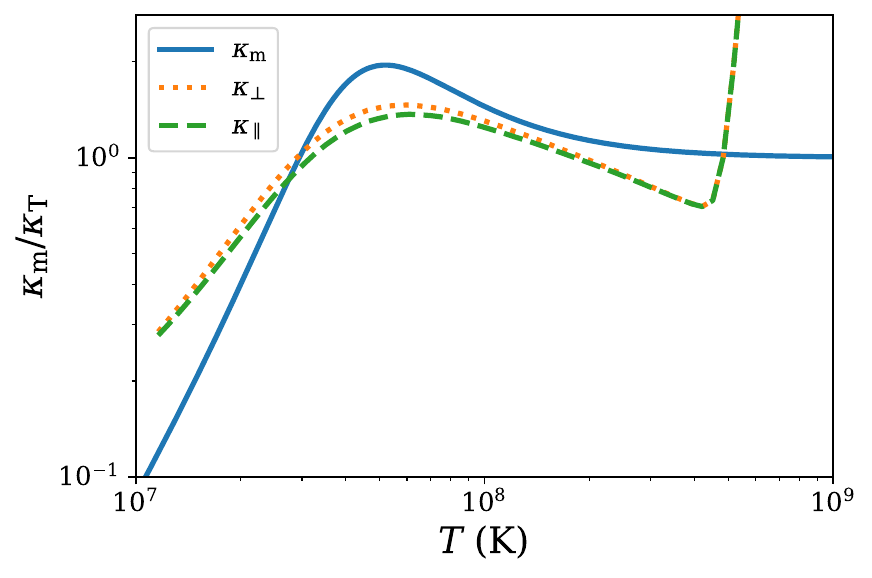}
\caption{Comparison between our magnetic opacity (solid line) and the perpendicular (dotted line) and parallel (dashed line) opacities of \citet{Suleimanov2022}, assuming $B=1\times10^{12}$~G and $\rho = 1~$g~cm$^{-3}$.  The very sharp increase in opacity at temperatures above $4\times10^8$~K is due to the onset of electron-positron pair production.}
\label{fig:suleimanov_comparison}
\end{figure}

We have shown in \autoref{subsec:opacity_instability} that instabilities can be present in accretion columns where the magnetic opacity is below the peak and therefore increases with temperature.  After we completed our simulations, new
opacity calculations by \citet{Suleimanov2022} were published.  These demonstrate that there is an even sharper increase of opacity with temperature when the medium is hot enough that pair production becomes significant (\autoref{fig:suleimanov_comparison}).  If this regime can exist within the column, we suspect that it will be a site of strong instability.  However, it may be that this opacity feature acts as a wall that cannot be reached in a real accretion column, because it may limit the temperature at the base of the column.  Once the accretion rate becomes high enough and the bottom of the accretion column starts to enter into the pair production regime, the strong opacity boost may provide additional radiation support and cause the column to expand.  This expansion can then cool the accretion column and bring the base temperature below the pair production value via the extra sideways emission.  However, similar to what we have found in the column oscillation, this expansion might then overcool the bottom of the column, contracting the column structure and heating the base back towards pair production.  This again suggests a strong destabilizing mechanism at high accretion rates and strong magnetic fields, in which a steady-state column simply cannot exist.  It would be interesting to simulate this in future, and  we are currently working on an algorithm for angle and polarization-dependent opacities for use in \textsc{Athena++}.  It would also be worthwhile to search for observational evidence of enhanced variability in this regime.

\subsection{Observational Significance}

One dimensional model fits to the observed pulse profiles of X-ray pulsars often require a mixture of fan and pencil-beam emission geometries, particularly at intermediate luminosities (e.g. \citealt{klo08,bec12}).  As shown in \autoref{fig:light_curve_ratio}, our high accretion rate simulations Highacc4 and Highacc6 both show almost pure ($>90$~percent) fan-beam emission.  However, in our lower accretion rate simulations, only Lowacc2 exhibits an almost pure fan beam emission pattern.  Lower magnetic field strengths instead produce a strongly oscillating column with an emission pattern that can vary between approximately half fan and pencil-beam and full fan-beam.  The time-average of these oscillations still result in $>80$~percent fan-beam, but this may be a contributing factor to the need for some pencil-beam emission for intermediate luminosities.  At the highest magnetic field strengths (Lowacc4 and Lowacc6), the beam patterns are fairly steady in time, with roughly 40 percent fan-beam in the case of Lowacc4, due largely to the fact that the vertical and horizontal projections of the photosphere are comparable in size. Lowacc6 is almost a hot spot, and emits only only 14 percent of its luminosity in a fan-beam.
It would therefore be interesting to explore observationally how the emission geometry depends not only on luminosity but also on magnetic field strength inferred from cyclotron lines.  That geometry can be strongly dependent on observed photon energy \citep{iwa19}, so post-processing of simulations such as ours would also be a useful way of confronting the observational constraints.

Accretion columns are observed to exhibit an inverse correlation between the luminosity and the magnetic field strength as measured by the energy of cyclotron lines \citep{Tsygankov2010,dor17,sta19,mus22}.  One possible explanation for this is that higher luminosities correspond to taller columns which then sample weaker magnetic fields in a diverging field geometry (but see \citealt{pou13} for an alternative explanation).  In simulations Lowacc4 and Highacc4, we increase the accretion rate of the column in the same magnetic field, and the column is taller.  Although we use a uniform, vertical magnetic field in the simulations, we can crudely estimate the corresponding decrease in field strength in a more realistic dipole geometry.  The luminosity increases by a factor of $\sim 23.59$, while the magnetic field strength at half the height of the column would decrease by a factor of $\sim 0.82$.  Similarly, when comparing simulations Lowacc6 and Highacc6, the luminosity increases by a factor of $\sim 35.07$ and the magnetic field would decrease by a factor of $\sim 0.81$.  These numbers are actually very close to the observed behavior of V~0332+53 \citep{Tsygankov2010,dor17}, where the luminosity increases by a factor of $\sim 20$ and the observed magnetic field strength decreases by a factor of $\sim 0.8$.  However, this source is inferred to have a surface field strength that is slightly weaker ($2.6\times10^{12}$~G) than these two pairs of simulations.  Moreover, as we have a uniform magnetic field in our simulations, the magnetic opacities may differ at higher altitudes. However, this difference is unlikely to significantly impact our results since the post-shock gas temperature should be sufficiently high so that the opacity is near the Thomson regime.  In addition, as our accretion columns are relatively short ($\sim 0.1$ neutron star radius), this crude estimation might still be reasonable.  Nevertheless, future simulations must incorporate the variation of magnetic field with height for a more accurate evaluation of this inverse correlation.

The QPOs that arise from the vertically shock oscillations are at frequencies in excess of $\simeq5$~kHz for the simulations we have presented in this paper, and this may prove challenging to observe directly with existing X-ray facilities.  However, the opacity-driven instability at high magnetic field strengths grows on much longer time scales $\simeq0.01$~s.  It is unclear how this will saturate and manifest in the lightcurve, as we were unable to run our simulations for long enough before the effective bottom boundary condition failed.  Even so, it suggests the possibility of longer time-scale variability that may be more easily observable for high magnetic field X-ray pulsars, whose fields are observed to extend up to as high as $6.6\times10^{12}$~G \citep{yam14,sta19}.  Further investigation of this instability with future simulations is therefore warranted.

\section{Conclusions}
\label{sec:conclusions}

We have extended our Cartesian simulations of \citet{Paper2} to incorporate polarization-averaged, temperature-dependent magnetic scattering opacities.  These can dramatically affect both the dynamics and the time-averaged structure of the accretion column.  For weak magnetic fields ($\simeq10^{11}$~G, simulation Lowacc01), the opacities inside the sinking zone are close to Thomson, and the column dynamics is very similar to what we found in \citet{Paper2}, the main difference being that coherent pre-shocks in the free-fall zone are largely absent because the opacity in that cold infalling material is much less than Thomson.  For higher magnetic field strengths, magnetic opacities produce much more significant differences, a result that is perhaps not surprising given that neutron star columns are supported against gravity by radiation pressure.

Increasing the magnetic field strength increases the temperature at which the opacity peaks, and also increases the width of that peak in temperature space, both effects scaling directly with the magnetic field strength.  At fixed accretion rate, increasing the magnetic field strength generally increases the post-shock opacity as an approximately fixed temperature jump across the shock is less and less able to climb over the opacity peak.  Despite the fact that the opacity in the interior of the column is on average larger, the time-averaged column height is reduced.  This is partly because a larger fraction of the immediate post-shock accretion power is radiated away rather than advecting into the higher opacity post-shock regions, and partly because more accretion power flows into the neutron star at the base of the column.

Again at fixed accretion rate, the taller columns at weaker magnetic field strengths $10^{11}-10^{12}$~G exhibit strong vertical oscillations.
While entropy waves (slow diffusion photon bubbles) are clearly present as horizontally-inward propagating waves, these oscillations are actually a result of global thermal imbalance, with over-cooling at maximum vertical extension and over-heating at minimum vertical extension.  This is the same mechanism that produced oscillations in our previous Thomson scattering simulations, both in Cartesian geometry \citep{Paper2} and split monopole geometry \citep{Paper3}.  The amplitude of these oscillations is large enough to cause the angular distribution of emitted radiation to oscillate from $50-80$~percent fan beam at minimum vertical extent to 100 percent fanbeam.  This might be a contributing factor to the need for 1D models to sometimes require a mixture of pencil beam and fan beam to explain observed light curves (e.g. \citealt{klo08,bec12,iwa19}).  The amplitude of these oscillations is reduced and the frequency is increased as the field strength increases at fixed accretion rate.  The angular distribution of emitted radiation is almost entirely fan beam until the field strength gets high enough that the column height is short enough to transition into almost a pure ($\sim85$ percent) pencil beam emitting hot spot (this happens at $6\times10^{12}$~G at the low accretion rates considered here).
Increasing the accretion rate at these higher magnetic field strengths restores the height of the column and the nearly 100~percent fan beam emission, but the oscillation amplitude remains quite low.  The mere fact that higher accretion rate results in taller columns may contribute to explaining the observed inverse correlation of cyclotron line energy with accretion rate during supercritical phases of accretion \citep{tsy06,jai16,sta19}.  But our simulations would also predict that the critical accretion rate separating hot spots (pencil beam emission, positive correlation between cyclotron energy and accretion rate) and columns (fan beam emission, negative correlation between cyclotron energy and accretion rate) should itself be a function of magnetic field strength across difference sources.

We have identified a new instability in the column that exists when the field strength is high enough that the opacity within the column increases inward with increasing temperature.  In the two high accretion rate simulations here, this instability grows on a time scale of $\sim0.01$~s.  While we were not able to fully investigate the nonlinear outcome of this instability due to numerical issues, it may contribute to enhanced variability on this time scale for high magnetic field neutron stars.

\section*{Acknowledgements}

We thank Ilaria Caiazzo, Jeremy Heyl, Matthew Middleton, Ekaterina Sokolova-Lapa, and Joern Wilms for useful conversations.  This work was supported in part by NASA Astrophysics Theory Program grant 80NSSC20K0525.  Computational facilities used in this research were purchased with funds from the National Science Foundation (CNS-1725797) and administered by the Center for Scientific Computing (CSC). The CSC is supported by the California NanoSystems Institute and the Materials Research Science and Engineering Center (MRSEC; NSF DMR 1720256) at UC Santa Barbara. The Center for Computational Astrophysics at the Flatiron Institute is supported by the Simons Foundation.

%%%%%%%%%%%%%%%%%%%%%%%%%%%%%%%%%%%%%%%%%%%%%%%%%%
\section*{Data Availability}

All the simulation data reported here is available upon request to the authors.



\bibliographystyle{mnras}
\bibliography{references} % if your bibtex file is called references.bib
%%%%%%%%%%%%%%%%%%%%%%%%%%%%%%%%%%%%%%%%%%%%%%%%%%

%%%%%%%%%%%%%%%%% APPENDICES %%%%%%%%%%%%%%%%%%%%%

\appendix

\section{Implementation of Angle, Polarization, and Frequency-Averaged Magnetic Scattering Opacities}
\label{sec:magnetic_opacity_derivation}

The scattering opacities in a neutron star accretion column depend strongly on both photon frequency, propagation angle with respect to the magnetic field, and polarization.  A multifrequency group method for isotropic absorption and emission has now been implemented in \textsc{Athena++} \citep{jia22}, and generalizing this to include angle-dependence and polarization would be worthwhile.  In the meantime, the simulations in this paper use the implementation of frequency-integrated radiation transfer in \textsc{Athena++} \citep{jiang2014,Jiang2021}.  Here we describe the angle, polarization, and frequency-averaged scattering opacities that we use.

We neglect quantum effects and treat photon-electron scattering classically, assuming a cold electron-ion plasma with infinite ion inertia and negligible plasma frequencies so that dispersive effects are negligible.  The full Mueller matrix for electron scattering in a uniform magnetic field under these conditions has been derived by \citet{chou1986}, and maps the Stokes parameters of the incoming to the outgoing radiation field.  We simplify the problem here by azimuthally averaging around the direction of the magnetic field \citep{caiazzo2021} and neglecting circular polarization.  Switching from Stokes parameters $I_\nu$ and $Q_\nu$ to
$O$ and $X$ mode intensities defined by $I_{O\nu}=(I_\nu+Q_\nu)/2$ and $I_{X\nu}=(I_\nu-Q_\nu)/2$, respectively, we can then derive the following scattering coefficients for $X$-modes,
\begin{equation}
    \chi_{X\nu}^{}=n_{\rm e}\sigma_{\rm T}\frac{\omega^2(\omega^2+\omega_{\rm ce}^2)}{(\omega^2-\omega_{\rm ce}^2)^2}\equiv fn_{\rm e}\sigma_{\rm T},
\end{equation}
$O$-mode radiation propagating perpendicular to the magnetic field,
\begin{equation}
    \chi_{O\nu}^\perp=n_{\rm e}\sigma_{\rm T}\left(\frac{4+f}{5}\right),
\end{equation}
and $O$-mode radiation propagating parallel to the magnetic field,
\begin{equation}
    \chi_{O\nu}^\parallel=n_{\rm e}\sigma_{\rm T}\left(\frac{2+3f}{5}\right).
\end{equation}
Here $\omega=2\pi\nu$ is the angular photon frequency, and $\omega_{\rm ce}=eB/(m_{\rm e}c)$ is the electron cyclotron angular frequency.
These expressions are identical to those in equation (45) of \citet{arons1987}, except that they used an approximation where $f$ is unity for $\omega>\omega_{\rm ce}$ and $\omega^2/\omega_{\rm ce}^2$ for $\omega<\omega_{\rm ce}$.  The exact (albeit classical) expression that we use here retains the enhancement of scattering near the cyclotron resonance.

Provided mode exchange is relatively efficient, the angle-averaged mean intensities in both polarization modes will be approximately equal \citep{arons1987}.  The polarization-averaged opacities for diffusion are then
\begin{equation}
\chi_{\nu}^\perp=\frac{2\chi_{O\nu}^\perp\chi_{X\nu}^{}}{\chi_{O\nu}^\perp+\chi_{X\nu}^{}}=\frac{f(4+f)}{2+3f}n_{\rm e}\sigma_{\rm T}
\end{equation}
and
\begin{equation}
\chi_{\nu}^\parallel=\frac{2\chi_{O\nu}^\parallel\chi_{X\nu}^{}}{\chi_{O\nu}^\parallel+\chi_{X\nu}^{}}=\frac{f(2+3f)}{1+4f}n_{\rm e}\sigma_{\rm T},
\end{equation}
in agreement with equation (50) of \citet{arons1987}.

We neglect finite photon chemical potential effects, and compute blackbody Rosseland means:
\begin{align}
    \chi_{{\rm R}\perp} &= n_{\rm e}\sigma_{\rm T}\frac{\int_0^\infty dx\frac{x^4e^x}
    {(e^x-1)^2}}{\int_0^\infty dx\frac{x^4e^x(2+3f)}{f(4+f)(e^x-1)^2}}
    \label{eq:perpross}
    \\
    &=n_{\rm e}
    \sigma_{\rm T}\times
    \begin{cases}
     1& \mbox{for } x_{\rm ce}\rightarrow0\\
    \dfrac{8\pi^2}{5x_{\rm ce}^2} & \mbox{for } x_{\rm ce}\rightarrow\infty
    \end{cases}
    \quad, 
\end{align}
and
\begin{align}
    \chi_{{\rm R}\parallel}&=n_{\rm e}\sigma_{\rm T}\frac{\int_0^\infty dx\frac{x^4e^x}
    {(e^x-1)^2}}{\int_0^\infty dx\frac{x^4e^x(1+4f)}{f(2+3f)(e^x-1)^2}}
    \label{eq:parross}
    \\
    &=n_{\rm e}
    \sigma_{\rm T}\times
    \begin{cases}
     1& \mbox{for } x_{\rm ce}\rightarrow0\\
    \dfrac{8\pi^2}{5x_{\rm ce}^2} & \mbox{for } x_{\rm ce}\rightarrow\infty
    \end{cases}
    \quad,
\end{align}
where $x\equiv h\nu/(kT)$ and $x_{\rm ce}=\hbar\omega_{\rm ce}/(kT)$.

To see how to implement these opacities, consider for simplicity a static medium.  The Rosseland mean opacities can then be incorporated into a frequency-integrated, polarization-averaged transfer equation as follows (cf. equation (6) of \citealt{Jiang2021}):
\begin{multline}    
\frac{1}{c}\frac{\partial I}{\partial t}+
\hat{\bf n}\cdot\nabla I
=\chi_{\rm Pa}\left(\frac{acT^4}{4\pi}-J\right)+\chi_{\rm Ra}(J-I)
\\
+\left[\frac{1}{2}(3\chi_{{\rm R}\perp}
-\chi_{{\rm R}\parallel})+
\frac{5}{2}(\chi_{{\rm R}\parallel}-\chi_{{\rm R}\perp})\cos^2\theta\right]
(J-I)
\quad. 
\label{eq:angletransfer}
\end{multline}
Here $\chi_{\rm Pa}$ and $\chi_{\rm Ra}$ are the Planck and Rosseland mean absorption coefficients for true absorption processes.  This equation automatically gives the correct zeroth and first moment equations for radiation energy density and flux, provided we assume nearly isotropic radiation closures on the second and third angular moments:  $K_{zz}=J/3$ and $Q_{izz}=(H_i+2\delta_{iz}H_z)/5$, where $J$ and $H_i$ are the zeroth and first angular moments, respectively.  The resulting moment equations are then
\begin{equation}
    \frac{1}{c}\frac{\partial J}{\partial t}+\nabla\cdot{\boldsymbol{H}}=\chi_{\mathrm{Pa}}\left(\frac{acT^4}{4\pi}-J\right),
\end{equation}
\begin{equation}
    \frac{1}{c}\frac{\partial H_i}{\partial t}+\frac{1}{3}\nabla_iJ=-(\chi_{\mathrm{Ra}}+\chi_{\mathrm{R}\perp})H_i-(\chi_{\mathrm{R}\parallel}-\chi_{\mathrm{R}\perp})\delta_{iz}H_z,
\end{equation}
exactly as required.  While we have worked in a static medium here, this is sufficient for the numerical radiation MHD scheme used in \textsc{Athena++}, which computes the source terms of the transfer equation in the local fluid rest frame and then Lorentz transforms them into the lab frame (see \citealt{Jiang2021} for details).

\begin{figure}
  \centering
  \includegraphics[width=\columnwidth]{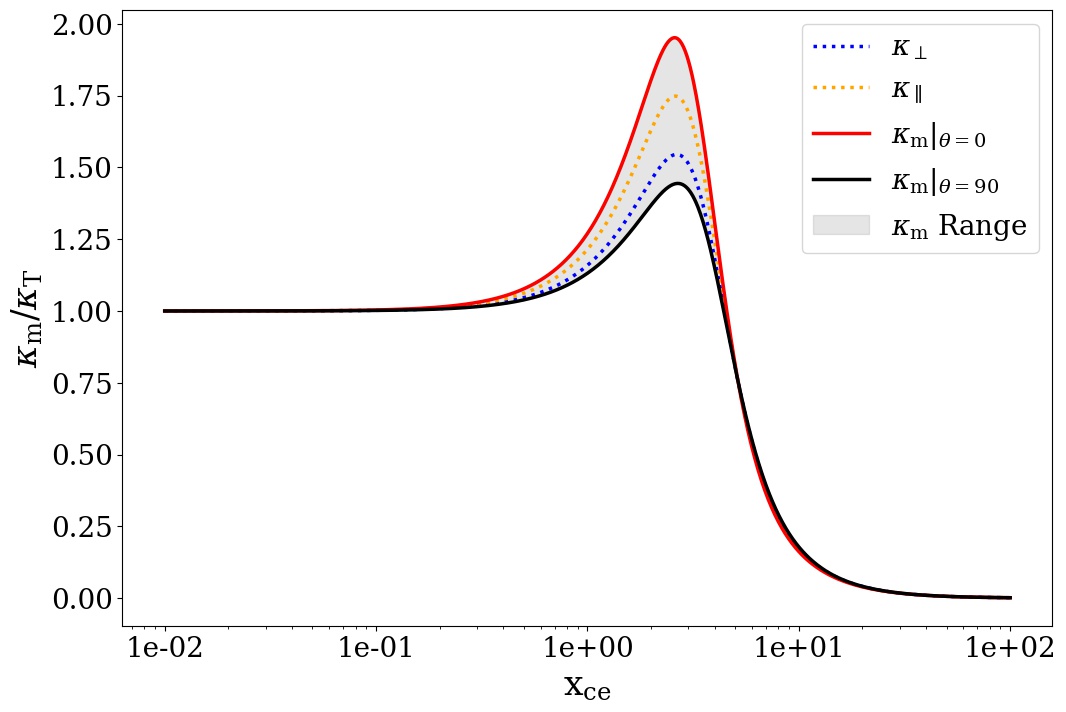}
  \caption{Behavior of the Rosseland scattering opacity, scaled with Thomson, as a function of $x_{\rm ce}=\hbar\omega_{\rm ce}/kT$.  Plotted are the perpendicular ($\kappa_{\mathrm{s},\perp}$), parallel ($\kappa_{\mathrm{s},\parallel}$), and the combination $0.5(3\kappa_{\mathrm{s},\perp}-\kappa_{\mathrm{s},\parallel})+2.5(\kappa_{\mathrm{s},\parallel}-\kappa_{\mathrm{s},\perp})\cos^2\theta$ for two propagation directions $\theta$. 
  }
  \label{fig:opacitylinear}
\end{figure}

\autoref{fig:opacitylinear} shows the behavior of the perpendicular and parallel Rosseland opacities ($\kappa_{\mathrm{s}}=\chi_{\mathrm{R}}/\rho$) from equations \autoref{eq:perpross} and \autoref{eq:parross} as a function of $x_{\rm ce}$, as well as the combination in the second line of equation \autoref{eq:angletransfer} for different angles of propagation.  All these curves are the same except near $x_{\rm ce}=1$, and we therefore simplify our transfer equation still further by neglecting the angle dependence in equation \autoref{eq:angletransfer}.  We do this by simply replacing $\cos^2\theta$ with unity, as that produces the largest opacity near $x_{\rm ce}=1$, in order to partly account for cyclotron resonance which we have neglected in this paper.  \autoref{fig:suleimanov_comparison} compares our prescription with the Rosseland mean opacities computed by \citet{Suleimanov2022}, and the agreement is reasonably good below the temperature $3\times10^8$~K where pair production becomes significant.  (We do not reach such temperatures in any of the simulations presented in this paper.)

To summarize, the frequency-integrated, polarization and angle-averaged magnetic scattering absorption coefficient that we actually use in the fluid rest frame is $2\chi_{{\rm R}\parallel}-\chi_{{\rm R}\perp}$, where $\chi_{{\rm R}\perp}$ and $\chi_{{\rm R}\parallel}$ are given by equations (\ref{eq:perpross}) and (\ref{eq:parross}), respectively.

\section{Photon Bubble Dispersion Relation With Variable Opacity}
\label{appendix:pbi_mag_opacity}

\citet{bla03} derived the dispersion relation for photon bubbles, including the effects of density and temperature-dependent opacities, in the short wavelength, rapid diffusion limit.  Here we allow for slow diffusion, generalizing the infinitely strong, vertical magnetic field analysis of \citet{Paper1} to incorporate variable Rosseland mean opacities $\kappa(\rho,T)$.  We define
\begin{equation}
    \Theta_{\rho}\equiv\frac{\partial\ln\kappa}{\partial\ln\rho}\,\,\,\,\,
    {\rm and}\,\,\,\,\,
    \Theta_{\rm T}\equiv\frac{\partial\ln\kappa}{\partial\ln T}.
\end{equation}
Then the only perturbation equation that changes from the linear analysis of a static, radiation pressure supported medium in Appendix A1 of \citet{Paper1} is their equation (A13) for the vertical radiative flux.  This now becomes
\begin{equation}
    \delta F_z=-\frac{c}{\kappa}\left[\frac{1}{\rho_0}\frac{\partial\delta P}{\partial z}+\frac{g}{\rho_0}(1+\Theta_\rho)\delta\rho+\frac{g}{4P_0}\Theta_T\delta P\right].
\end{equation}
Carrying through the analysis of \citet{Paper1}, assuming a spacetime dependence for the perturbations $\propto\exp[i(k_x x+\int k_zdz-\omega t)]$, we arrive at a cubic dispersion relation for the frequencies similar to equation (A18) of that paper:
\begin{eqnarray}
0&=&\omega^3+\omega^2\left[ik^2\left(\frac{c}{3\kappa\rho_0}+V\right)
+\frac{cg\Theta_Tk_z}{9\kappa\rho_0 c_{\rm r}^2}\right]\cr
&&+\omega\left(-k_z^2c_{\rm r}^2-\frac{k^4cV}{3\kappa\rho_0}
+\frac{ik^2Vcg\Theta_T k_z}{9\kappa\rho_0 c_{\rm r}^2}\right)\cr
&&+\left(-\frac{cg}{3\kappa\rho_0}k_x^2k_z+\frac{ik_z^2g^2c
\Theta_T}{9\kappa\rho_0 c_{\rm r}^2}+\frac{cgk_z^3\Theta_\rho}{3\kappa\rho_0}\right).
\label{eq:thetadisp}
\end{eqnarray}

The viscous ($V$) terms generally set the short wavelength cutoff scale when finite gas pressure effects are negligible.  Neglecting these terms and taking the short wavelength $k\rightarrow\infty$ limit, we find that there is a mode given by
\begin{equation}
    \omega^2=-\frac{igk_z}{k^2}(k_x^2-k_z^2\Theta_\rho).
\end{equation}
This agrees exactly with equation (98) of \citet{bla03} for a vertical magnetic field, and generalizes the dispersion relation of \citet{gam98} to incorporate the effects of a variable opacity.  Temperature fluctuations are smoothed out by rapid diffusion at short wavelengths, which is why $\Theta_{\rm T}$ does not appear in this limit.  Indeed, note that in the last three terms of equation (\ref{eq:thetadisp}), the $\Theta_T$ term is at one order of $k$ lower than the other two terms.

The magnetic opacities that we have considered in this paper have no density-dependence:  $\Theta_\rho=0$.  In addition, the photon bubble instability requires $k_x\ne0$.  If we eliminate photon bubbles by considering purely vertically propagating modes, and adopt $\Theta_\rho=0$, we find a short wavelength mode with frequency given by
\begin{equation}
    \omega^2+i\frac{3\kappa\rho_0c_{\rm r}^2}{c}\omega+\frac{g^2}{3c_{\rm r}^2}\Theta_T=0.
    \label{eq:mag_instability_approx}
\end{equation}
One of the roots of this equation is always unstable if $\Theta_T>0$.  Because it is entirely vertical, the magnetic field plays no role here except in determining the temperature dependence of the opacity, and the instability is fundamentally hydrodynamic in nature.
%%%%%%%%%%%%%%%%%%%%%%%%%%%%%%%%%%%%%%%%%%%%%%%%%%

% Don't change these lines
\bsp	% typesetting comment
\label{lastpage}
\end{CJK*}
\end{document}